%% file: main.tex
\documentclass{article}

\usepackage{longtable}
\usepackage{microtype}
\usepackage{graphicx}
\usepackage{subcaption}
\usepackage{booktabs} 
\usepackage{float}
\usepackage{placeins} 
\usepackage[normalem]{ulem}
\usepackage{enumitem}


\usepackage{hyperref}

\usepackage[preprint]{icml2026}

\usepackage{amsmath}
\usepackage{amssymb}
\usepackage{mathtools}
\usepackage{amsthm}
\usepackage{tikz}
\usetikzlibrary{
    positioning, 
    calc, 
    arrows.meta, 
    shadows.blur, 
    shapes.geometric, 
    3d, 
    backgrounds,
    fit,
    decorations,
    patterns
}

\definecolor{pnasBlue}{RGB}{0, 114, 178}     
\definecolor{pnasLightBlue}{RGB}{86, 180, 233} 
\definecolor{pnasRed}{RGB}{213, 94, 0}       
\definecolor{pnasOrange}{RGB}{230, 159, 0}   
\definecolor{pnasGreen}{RGB}{0, 158, 115}    
\definecolor{pnasGray}{RGB}{240, 240, 240}   
\definecolor{pnasDarkGray}{RGB}{80, 80, 80}  

\usepackage[capitalize,noabbrev]{cleveref}

\theoremstyle{plain}

\theoremstyle{definition}

\theoremstyle{remark}

\newcommand{\bx}{\boldsymbol{x}}

\usepackage[textsize=tiny]{todonotes}

\begin{document}

\twocolumn[
\icmltitle{Hybrid operator learning of wave scattering maps in high-contrast media}



\icmlsetsymbol{equal}{*}

\begin{icmlauthorlist}
  \icmlauthor{Advait Balaji}{equal,comp}
  \icmlauthor{Trevor Teolis}{equal,yyy}
  \icmlauthor{S. David Mis}{equal,yyy}
  \icmlauthor{Jose Antonio Lara Benitez}{yyy}
  \icmlauthor{Chao Wang}{comp}
  \icmlauthor{Maarten V. de Hoop}{yyy}
\end{icmlauthorlist}

\icmlaffiliation{yyy}{Department of Computational and Applied Mathematics, Rice University, Houston, TX, USA}
\icmlaffiliation{comp}{Occidental Petroleum Corporation, Houston, TX, USA}

\icmlcorrespondingauthor{Advait Balaji}{Advait\_Balaji@oxy.com}
\icmlcorrespondingauthor{Maarten V. de Hoop}{mvdh@rice.edu}

\icmlkeywords{operator learning, transformer, Helmholtz, Lippmann Schwinger}

\vskip 0.3in
]



\printAffiliationsAndNotice{\icmlEqualContribution} 

\input{00_Abstract}
\input{01_Introduction}

\input{04_Prior-work}

\input{02_Background}

\input{03_Architectures}

\input{05_Experiments}

\input{06_Discussion}

\section*{Impact Statement}

We present a hybrid neural surrogate for high-frequency Helmholtz wave propagation in high-contrast media. 
Such surrogates can lower the cost of forward modeling and accelerate research workflows in seismic imaging and inversion, including applications like subsurface monitoring, CO$_2$ sequestration assessment, and geothermal exploration. The same capabilities could also be applied toward hydrocarbon exploration or other dual-use sensing tasks, and training deep models consumes energy and compute resources.

\bibliography{refs}
\bibliographystyle{icml2026}

\clearpage

\input{07_appendix}

\end{document}

%% file: 00_Abstract.tex
\begin{abstract}

Surrogate modeling of wave propagation and scattering (i.e. the wave speed and source to wave field map) in heterogeneous media has significant potential in applications such as seismic imaging and inversion.  
High-contrast settings, such as subsurface models with salt bodies, exhibit strong scattering and phase sensitivity that challenge existing neural operators.  
We propose a hybrid architecture that decomposes the scattering operator into two separate contributions: a smooth background propagation and a high-contrast scattering correction.  
The smooth component is learned with a Fourier Neural Operator (FNO), which produces globally coupled feature tokens encoding background wave propagation; these tokens are then passed to a vision transformer, where attention is used to model the high-contrast scattering correction dominated by strong, spatial interactions.   
Evaluated on high-frequency Helmholtz problems with strong contrasts, the hybrid model achieves substantially improved phase and amplitude accuracy compared to standalone FNOs or transformers, with favorable accuracy–parameter scaling.  

\end{abstract}

%% file: 01_Introduction.tex
\section{Introduction}
\label{sec:intro}

Forward modeling of wave propagation and scattering through heterogeneous media is fundamental to seismic inverse (boundary) problems---a central tool for subsurface characterization in applications ranging from hydrocarbon exploration, reservoir monitoring, CO$_2$ sequestration, and geothermal energy \cite{Sheriff1995ExplorationSeismology, Yilmaz2001SeismicDataAnalysis, virieux2009overview, lumley2001, arts2004sleipner, Chadwick2009BestPractice, Schmelzbach2016GeothermalImaging}.
These problems can be formulated using the time-harmonic Helmholtz equation \cite{pratt1999, Pratt1990CrossholeTomography} or the equivalent Lippmann--Schwinger integral equation \cite{Clayton1981BornWKBJ, deHoop2000WavefieldReciprocity}, and form the backbone of algorithms like Full Waveform Inversion (FWI) \cite{tarantola1984, virieux2009overview, Fichtner2011Book}. 

Traditional solvers like Finite-Difference Methods (FDM) \cite{Virieux1984SH, Moczo2002HeterogeneousFD, Robertsson1994ViscoelasticFD, Cheng2021qPPropagators} and Finite-Element Methods (FEM) \cite{Marfurt1984FDvsFEMAccuracy, Padovani1994FEMSeismic, Koketsu2004VoxelFEM, florian2025} accurately resolve complex wave phenomena but require a full forward solve each time the wavespeed model is updated, leading to prohibitive computational cost in inverse problems \cite{virieux2009overview, Tromp2005BananaDoughnut, Fichtner2006AdjointTheory}. Data-driven surrogate models offer faster inference once trained, though learning the solution operator remains challenging. Candidate architectures include Physics-Informed Neural Networks (PINNs) \cite{Raissi2019PINNs, Song2022PINNsVTI, Huang2024PINNsFormer}, which incorporate the PDE through a penalty term but require training for each model \cite{Wang2021WhenPINNsFail, Krishnapriyan2021PINNFailureModes}, and neural operators such as Fourier Neural Operators (FNOs) \cite{li2020fourier, lara_benitez2024ood, liu2024wavebench, Yin2023ImplicitForecasting}, which enable efficient, mesh-independent inference and have demonstrated strong performance for PDEs with smooth coefficients \cite{NO-Stuart-Kovachki-Anandkumar-et-al}.

This work builds on the analysis of out-of-distribution generalization for neural operators in \cite{lara_benitez2024ood}, shifting emphasis from smooth wavespeed to representation and modeling in high-contrast scattering regimes. Many geological settings involve large obstacles (such as salt bodies) which admit wavespeed contrasts that generate strong scattering in the wavefield, leading to a more challenging learning problem \cite{virieux2009overview}. In such high-contrast regimes, scattering induces strong geometry-dependent spatial interactions across the domain, falling outside the inductive bias of architectures designed for smoothly propagating fields. 
We argue that attention-based architectures are well-suited for these geometry-dependent interactions. 
For operator learning, the vision transformer is the architecture of choice. 
While originally designed for vision tasks such as object recognition and segmentation, recent architectures such as scOT have demonstrated strong performance as surrogate models for various PDEs \cite{Herde2024Poseidon}.

Nonetheless, a clear understanding of when and why transformer-based models outperform classical neural operators in challenging physical regimes remains limited. This work demonstrates that high-contrast wave scattering constitutes a distinct regime where transformer-based models outperform classical neural operators. 
Unlike smooth background propagation, these high-contrast regimes benefit from the global receptivity of self-attention mechanisms that can selectively adapt interaction patterns to complex scattering structure.

We introduce a hybrid architecture, which is driven by the physics of wave propagation and scattering.  
 We split the wavespeed into a global, smooth background and a sharp localized contrast. An FNO acts on the smooth background and generates globally coupled tokens for a vision transformer that acts on the sharp contrast. 
 One can view the hybrid design as an inductive bias for scattering problems with sharp obstacles.  Our main contributions are as follows: 

\begin{itemize}
    \item \textbf{Scattering-aware operator decomposition.}  
    We propose an initial splitting of the wavespeed model, inducing a decomposition of the Helmholtz forward map into a smooth background propagation operator and a high-contrast scattering correction. This formulation isolates the effects of localized wavespeed discontinuities into a separate operator-learning problem. The resulting decomposition yields a better-conditioned learning task and explicitly reflects the underlying physics of wave propagation and scattering.

    \item \textbf{Hybrid neural operator--transformer architecture.}  
    Building on this decomposition, we introduce a hybrid architecture that combines a neural operator (e.g. FNO) to model the smooth background field with a vision-transformer-based model (e.g. scOT) to learn the high-contrast scattering corrector. The transformer employs patch-based representations and shifted-window self-attention to efficiently capture the strong spatially dependent interactions induced by scattering, while maintaining linear complexity with respect to the number of patches.

    \item \textbf{Empirical validation on challenging Helmholtz benchmarks.}  
    We evaluate the proposed framework on high-frequency (40 Hz) Helmholtz problems with wavespeed contrasts and complex scattering phenomena. Across these benchmarks, the hybrid model consistently outperforms standalone FNOs and transformer baselines in accuracy, and exhibits favorable accuracy--parameter scaling in strongly scattering regimes.
\end{itemize}

%% file: 04_Prior-work.tex
\section{Prior work}

\label{sec:prior-work}

\subsection{Neural Operator Architectures}
Recent advances in neural operators (NOs) have enabled efficient surrogate modeling for PDEs. Fourier Neural Operators (FNOs) offer discretization-invariant inference on fixed Cartesian grids and perform well for smooth or weakly heterogeneous problems \cite{li2020fourier,Kovachki2021FNOTheory}, but struggle in wave propagation settings with strong contrasts \cite{liu2024wavebench}. This limitation aligns with recent theory showing that the effective rank and parameter complexity of kernel-based neural operators grow rapidly as the Sobolev regularity of the input–output map decreases \cite{kratsios2024mixture}. Several hybrid extensions address these challenges: Fourier-DeepONet \cite{Zhu2023FourierDeepONet} integrates FNOs within DeepONet for robustness in full waveform inversion, NSNO \cite{Chen2024NSNO} incorporates a Neumann-series–inspired scattering expansion with multiscale features to reduce high-frequency errors, and deep neural Helmholtz operators \cite{zou2024deep} and physics-guided generative neural operators \cite{cheng2025seismicgno} target large-scale wave propagation and distributions of scattered fields under physical constraints.

Complementing these hybrid and generative formulations,  Convolutional Neural Operators (CNO) have been adopted to enhance spatial locality and boundary handling~\citep{raonic2023convolutional}. These structural advances are often paired with transfer learning to facilitate physical adaptation~\citep{wang2024transfer} or physics-informed training schemes~\citep{ma2025picno,song2024traveltime}. However, despite these improvements, neural operators continue to face significant limitations, including a pronounced low-frequency spectral bias~\citep{lehmann2024ffno,qin2024spectral} and poor generalization to unseen geological structures~\citep{ma2025picno}. Crucially for inversion, prediction errors on perturbed models propagate to yield noisy gradients and instability~\citep{huang2025physics}. Recent work addressing these specific stability issues via physics-based constraints---such as PDE penalty terms---has achieved significant error reductions~\citep{ma2025picno,cheng2025gno,huang2025physics}.
\citet{Alkhalifah2024-scattered-residual} learn the FWI residual with a neural operator in layered sedimentary media.

\begin{figure}
    \centering
    \begin{tikzpicture}[scale=1.2, >=latex, font=\small]

        \shade[top color=cyan!5, bottom color=cyan!12] (0,3.0) rectangle (6,4);

        \fill[orange!5] (0,1.8) -- (0,3.0) -- (6,3.0) -- (6,2.0)
        .. controls (4,2.2) and (2,1.6) .. (0,1.8) -- cycle;

        \fill[brown!10] (0,0) -- (0,1.8)
        .. controls (2,1.6) and (4,2.2) .. (6,2.0) -- (6,0) -- cycle;

        \draw[gray!40, thin] (0,3.0) -- (6,3.0); 
        \draw[gray!40, thin] (0,1.8) .. controls (2,1.6) and (4,2.2) .. (6,2.0); 

        \begin{scope}
            \fill[red!85!black]
            plot [smooth cycle, tension=0.6] coordinates {
                    (3.8, 0.5)
                    (3.4, 1.2)
                    (3.3, 2.0)
                    (3.7, 2.4)
                    (4.0, 2.2)
                    (4.4, 2.5)
                    (4.4, 2.0)
                    (4.4, 1.0)
                    (4.2, 0.4)
                };

            \node[text=white, font=\bfseries, align=center, scale=0.9] at (4.0, 1.5) {Salt\\Body};
        \end{scope}

        \draw[very thick, black!80] (0,0) rectangle (6,4);

        \node[anchor=south west, text=black!80] at (0,4.05) {$\text{$\Gamma_{\mathrm{Free}}$}$};
        \node[anchor=north west, text=black!80] at (0,-0.05) {$\Gamma_{\mathrm{ABC}}$};
        \node[rotate=-90, anchor=south, text=black!80] at (6.05, 3.5) {$\Gamma_{\mathrm{ABC}}$};
        \node[rotate=90, anchor=south, text=black!80] at (-0.05, 3.5) {$\Gamma_{\mathrm{ABC}}$};

        \coordinate (S) at (3.2, 3.5);
        \fill[black] (S) circle (2pt);
        \node[above, text=black, yshift=2pt] at (S) {Source};

        \foreach \r in {0.3, 0.6, 0.9} {
                \draw[black!50, thick] (S) ++(210:\r) arc (210:330:\r);
            }

        \node[align=left, fill=white, draw=gray!30, inner sep=10pt, anchor=north] at (3, -0.5) {
            \textbf{Absorbing Boundary Condition:}\\
            $\displaystyle \partial_\nu p - i \frac{\omega}{c(\mathbf{x})} p = 0 \quad \text{on } \Gamma_{\mathrm{ABC}}$ 
        };

    \end{tikzpicture}
    \caption{Schematic of a heterogeneous wave domain containing a high-contrast obstacle (i.e. salt body).
    On $\Gamma_{\mathrm{ABC}}$ we impose an absorbing (Sommerfeld)
    boundary condition so outgoing waves exit without spurious reflections. $\Gamma_{\mathrm{Free}}$ is a free-surface boundary, with $p = 0$.}
    \label{fig:boundary-conditions}
\end{figure}
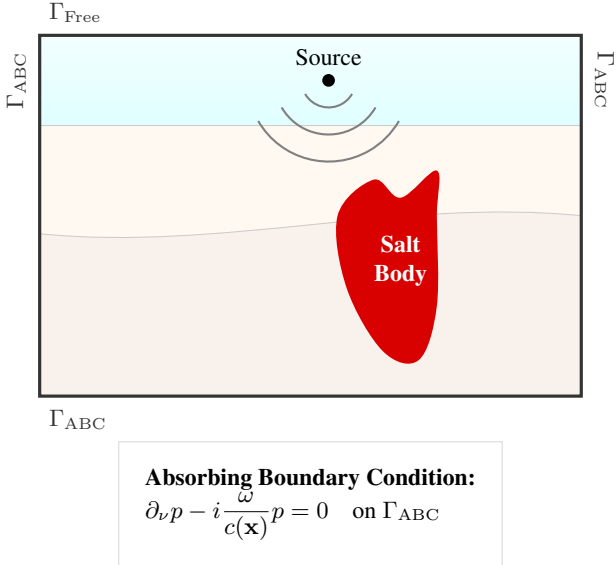

\subsection{Transformer-based PDE Solvers}
Recently, transformer-based approaches to operator learning have been developed as an alternative to neural operators.  While neural operators admit rigorous approximation-theoretic frameworks and universal approximation results, an analogous theoretical foundation for transformers in operator learning remains largely undeveloped. 

Operator transformers like OFormer \cite{Li2022TransformerPDE} demonstrated that self and cross-attention mechanisms
can be used to learn solution operators in a discretization-agnostic manner. 
Building on OFormer, Dong et al.~\cite{Dong2023GNOT} introduced the General Neural Operator Transformer (GNOT), combining operator-style attention with geometry-aware encodings for irregular discretizations. Related efforts such as Multiple Physics Pretraining (MPP) \cite{McCabe2024MultiplePhysics} and the Geometry-Aware Operator Transformer (GAOT) \cite{wen2026geometryawareoperatortransformer} improve generalization and accuracy on smooth PDEs and complex geometries, but do not address high-frequency scattering with strong wavespeed contrasts.

Adaptive Fourier Neural Operators (AFNOs) \cite{Guibas2021AdaptiveFN} repurpose FNO-style spectral mixing as a token-mixing mechanism for transformers, achieving quasi-linear complexity via block-diagonal channel mixing and shared frequency-domain MLPs. While AFNO is designed primarily as a transformer token mixer rather than a standalone PDE surrogate, it has been adopted in operator-learning settings such as the Denoising Operator Transformer (DPOT) \cite{Hao2024DPOT}, which replaces standard attention with spectral mixing to scale across PDE families. Relatedly, \cite{HUANG2024-LordNet} introduced LordNet, a fully convolutional Fourier-based surrogate that captures long-range dependencies through channel mixing but does not employ token-wise attention. 


%% file: 02_Background.tex
\section{Scattering problem}

\label{sec:helmholtz}

We consider the Helmholtz model of wave propagation.
Let $\omega$ denote angular frequency, and $\Omega \subset \mathbb{R}^2$ be a spatial rectangular domain. Consider the wavespeed $v(\bx)$  and time-harmonic pressure field $p(\bx,\omega)$ at location $\bx \in \Omega$. The pressure $p(\bx,\omega)$ is generated by a source term $s(\bx,\omega)$ and satisfies the Helmholtz equation, 
\begin{equation}\label{eq:helmholtz}
\begin{cases}
    \left[ \Delta + \frac{\omega^2}{v(\bx)^2} \right] p(\bx,\omega) = -s(\bx,\omega) & \bx \in \Omega, \\
    \partial_\nu p(\bx,\omega) - \frac{i\omega}{v(\bx)}\, & \bx \in \Gamma_{\mathrm{ABC}}, \\
    p(\bx, \omega) = 0 & \bx \in \Gamma_{\mathrm{Free}},
\end{cases}
\end{equation}
where $\Delta$ is the Laplace operator with respect to the spatial coordinates $\bx$, and $\partial_\nu$ is the normal derivative with respect to the boundary $\Gamma_{\mathrm{ABC}}$, which consists of three edges of the rectangular domain.  The remaining edge, $\Gamma_{\mathrm{Free}}$, is a free-surface boundary, with Dirichlet condition $p=0$. 
The boundary conditions are illustrated in Figure~\ref{fig:boundary-conditions}. 

Even for smooth wavespeeds $v(\bx)$, the Helmholtz operator exhibits phase evolution. In high-contrast media, however, variations in $v(\bx)$ give rise to strong scattering effects, including reflections, refractions, and multiple scattering, leading to spatially heterogeneous and geometry-dependent interaction patterns. Accurately modeling this regime requires capturing both smooth background propagation and contrast-driven scattering effects.

\subsection{Forward map}
We study the forward map associated with the Helmholtz equation, which maps a spatially varying wavespeed to the resulting pressure field.  Specifically, given a frequency $\omega$, the forward map 
\begin{equation*}
    \mathcal{F}: (s(\cdot, \omega), v) \mapsto p(\cdot, \omega)
\end{equation*}
assigns to each source $s$ and wavespeed $v$ the corresponding solution $p(\bx, \omega)$ given by the solution of \eqref{eq:helmholtz}.

\subsection{Scattering decomposition}

In heterogeneous media, the forward map $\mathcal{F}$ from wavespeed to pressure field admits a natural structural decomposition into a smoothly varying background response and a secondary contribution induced by strong contrasts, interfaces, or localized inclusions. The latter arises from the discrepancy between the true wave speed and its smoothed approximation and manifests itself as a residual field. This residual satisfies a wave equation with an effective source term that depends explicitly on the contrast source term that also depends on the background wavefield; when expressed in integral form, this yields a Lippmann–Schwinger representation. This formulation makes clear that the residual is not merely a correction in amplitude, but a mechanism for generating new wave phenomena through scattering.

We now formulate the Helmholtz equation in a way that exposes the structural decomposition of the scattering.  We decompose the wavespeed as 
\[
\delta m(\bx) = v^{-2}(\bx) - v_{\mathrm{bg}}^{-2}(\bx),
\] 
where $v_{\mathrm{bg}}(\bx)$ is a smoothed wavespeed obtained by applying a mollifying convolution to $v(\bx)$, see Appendix~\ref{app:salt_image_blur}.  An example of a high-contrast wavespeed decomposed into its smoothed and residual components is shown in Figure~\ref{fig:decomposition}.

Correspondingly, we decompose the pressure field
\[
    \delta p(\bx, \omega) = p(\bx, \omega) - p_{\mathrm{bg}}(\bx, \omega), 
\]
where $p$ is the solution to \eqref{eq:helmholtz} and $p_{\mathrm{bg}}(\bx, \omega)$ is the solution to the same equation corresponding to the smooth ``background'' wavespeed $v = v_{\mathrm{bg}}$. 
The equation for $\delta p$ is then given by 
\begin{equation}\label{eq:helmholtz-residual}
\begin{cases}
    \left[ \Delta + \frac{\omega^2}{v_{\mathrm{bg}}(\bx)^2} \right] \delta p(\bx,\omega) = \\-\omega^2 \delta m(\bx) \big(\delta p(\bx, \omega) + p_\mathrm{bg}(\bx, \omega)\big) & \bx \in \Omega, \\
    \partial_\nu \delta p(\bx,\omega) - \frac{i\omega}{v(\bx)}\, \delta p(\bx,\omega) = 0,
& \bx \in \Gamma_{\mathrm{ABC}}, \\
    \delta p(\bx, \omega) = 0 & \bx \in \Gamma_{\mathrm{Free}},
\end{cases}
\end{equation}
which can be solved with the method of Green's functions, giving rise to the Lippmann-Schwinger integral formulation, see Appendix~\ref{app:greens-function}.
Our emphasis is on the PDE level, which informs our operator-learning design.

The splitting of the wavefield into the smooth and residual part gives rise to an operator splitting of the forward map $\mathcal{F}$ into two distinct physical parts: (i) a smooth background map $\mathcal{F}_\mathrm{bg}$, and (ii) a high-contrast scattering corrector $\mathcal{F}_\mathrm{sc}$.
For a fixed frequency $\omega$, the background map,
\[
    \mathcal{F}_\mathrm{bg}: (s(\cdot, \omega), v_\mathrm{bg}) \mapsto  p_\mathrm{bg}(\cdot, \omega),
\]
assigns to a source $s(\cdot, \omega)$ and a smooth background wavespeed, $v_\mathrm{bg}$, the corresponding field $p_\mathrm{bg}$.  
In other words, $\mathcal{F}_\mathrm{bg}$ is the forward map restricted to smooth background fields.
Given the background pressure field $p_\mathrm{bg}(\cdot, \omega)$ obtained from $\mathcal{F}_\mathrm{bg}$, the high-contrast scattering corrector,
\[
    \mathcal{F}_\mathrm{sc}: (p_\mathrm{bg}(\cdot, \omega), \delta v) \mapsto \delta p ,
\]
assigns the $p_\mathrm{bg}$ and the residual  $\delta v$ to the pressure field perturbation $\delta p$ solving \eqref{eq:helmholtz-residual}.
We make the dependence on $\delta v$ instead of $\delta m$, suppressing its explicit dependence on $v_\mathrm{bg}$. 


The full forward map $\mathcal{F}: (s(\cdot, \omega), v) \mapsto p(\cdot, \omega)$ is then recovered by superposition:  
\[
    p(\bx, \omega) = \delta p(\bx, \omega) + p_\mathrm{bg}(\bx, \omega). 
\]


%% file: 03_Architectures.tex
\begin{figure*}[t]
\label{fig:decomposition}
\centering

\includegraphics[width=\textwidth,trim={5cm 6cm 6cm 1cm},clip]{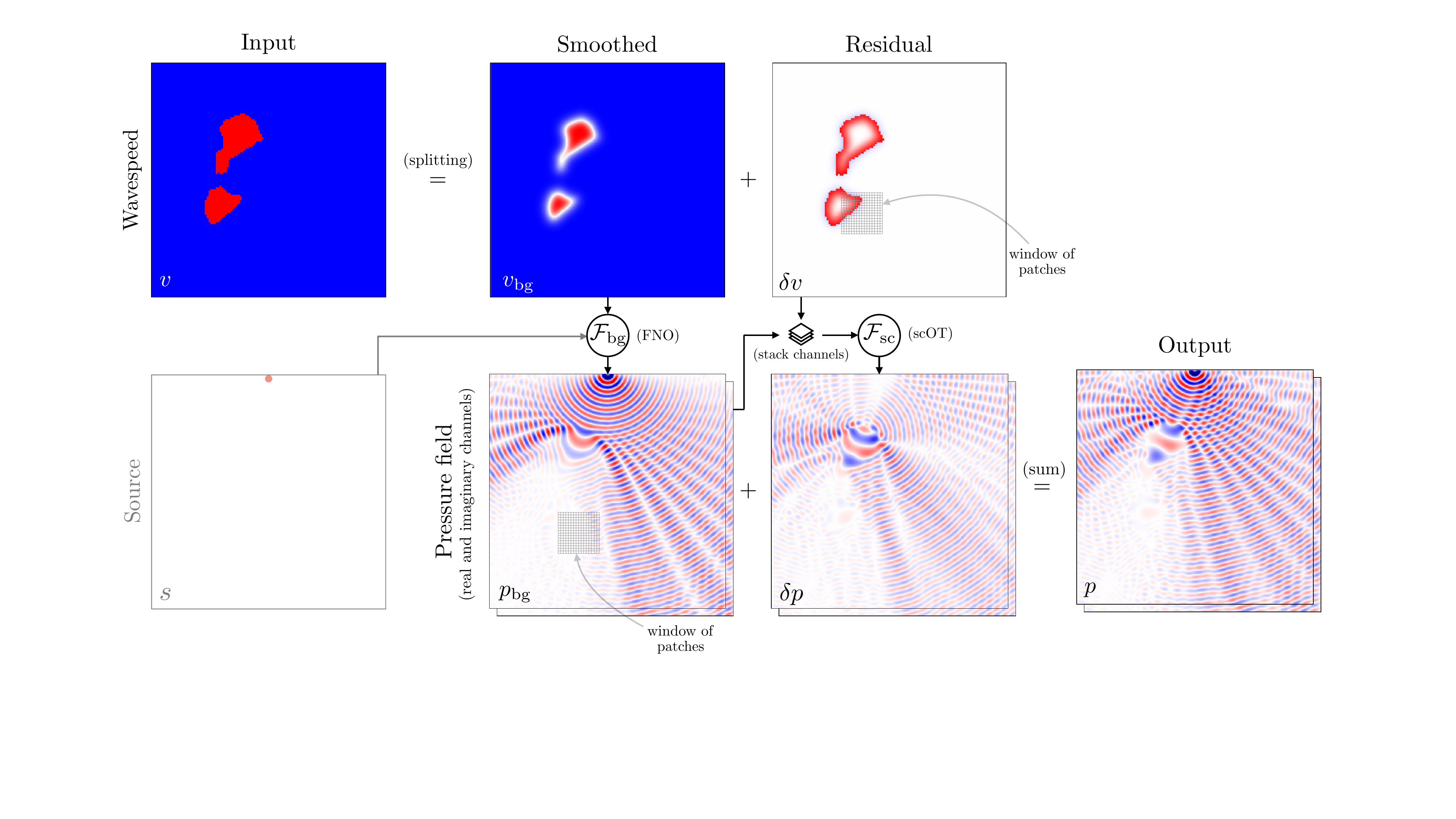}

\caption{Decomposition of the Helmholtz forward map into a smooth background propagation
$\mathcal F_{\mathrm{bg}}$ and a scattering correction $\mathcal F_{\mathrm{sc}}$. Our proposed hybrid model uses an FNO to learn the $F_{\mathrm{bg}}$ map from source $s$ and smoothed wavespeed $v_\text{bg}$ to the background pressure field $p_\text{bg}$, and a Vision Transformer to learn the $\mathcal F_{\mathrm{sc}}$ map from $p_\text{bg}$ and wavespeed residual $\delta v$ to the pressure residual $\delta p$. The final output is $p_\text{bg} + \delta p$. For the experiments in this work, we use a constant point source near the center of the free boundary; surrogate FNOs with varied sources for smooth wavespeed were investigated in \cite{lara_benitez2024ood}.}  
\label{fig:decomposition}
\end{figure*}

\section{Overview of Architecture}

\label{sec:architecture}

Casting the residual as a separate operator-learning problem has significant conceptual and practical advantages. It isolates the singular and spatially localized effects---such as reflections and diffractions---that are fundamentally distinct from the propagation and phase evolution governed by the smoothed background. Rather than requiring a single architecture to capture both regimes simultaneously, the decomposition aligns each component with a learning model suited to its mathematical character: neural operators efficiently approximate the mapping from smooth wavespeeds, while transformer-based architectures are suited to represent the interactions and complex dependencies inherent in the scattering-induced residual from local singular contrasts.  
This leads to our hybrid architecture, consisting of
\begin{itemize}
    \item an FNO that learns the smooth operator, $\mathcal{F}_{\mathrm{bg}}$, corresponding to the smooth background field $v_\mathrm{bg}(\bx)$, and
    \item a vision transformer (e.g. scOT) that learns the high-contrast corrector, $\mathcal{F}_{\mathrm{sc}}$, corresponding to the discontinuous background field $\delta v(\bx)$.
\end{itemize}

The following algorithm computes the forward map.
Given a fixed frequency $\omega$, a wavespeed $v$, and a source $s(\cdot, \omega)$, 
\begin{enumerate}[label=(\roman*)]
    \item Evaluate the smoothed $v_\mathrm{bg}$ by applying
    a mollifying convolution.
    \item Evaluate the contrast 
    $\delta v(\bx) = v(\bx) - v_\mathrm{bg}(\bx)$. 
    \item Using an FNO, predict the pressure field corresponding to the smooth wavespeed: $p_\mathrm{bg} = \mathcal{F}_\mathrm{bg}(v_\mathrm{bg})$.
    \item Using a vision-transformer, predict the pressure field corresponding to the contrast: $\delta p = \mathcal{F}_\mathrm{sc}(p_\mathrm{bg}, \delta v)$. 
    \item Recover the full pressure field $p = \delta p + p_\mathrm{bg}$. 
\end{enumerate}

Figure~\ref{fig:decomposition} illustrates the algorithm on representative example inputs. Source conditioning can be handled via a hypernetwork \cite{lara_benitez2024ood}, but here we use a fixed Gaussian source.


The design of our hybrid architecture is motivated by recent theoretical results on the approximation properties of neural operators. In particular, Kratsios et al.~\cite{kratsios2024mixture} show that the rank required to achieve a given accuracy increases sharply as the regularity of the input--output map decreases, highlighting fundamental challenges in low-regularity (i.e. high-contrast) regimes. This theory helps explain the empirical difficulty of learning operators with high-contrast inputs using a single neural operator. Consequently, while neural operators are well-suited for approximating the smooth background map $\mathcal{F}_{\mathrm{bg}}$, representing the high-contrast scattering operator $\mathcal{F}_{\mathrm{sc}}$ would require prohibitively large model capacity. The separation leads to a better-conditioned learning problem by reducing the disparity of scales and regularity that a single network would otherwise need to resolve. 

\subsection{Neural operator for smoothed wavespeed}

In our hybrid model, an FNO is used as a high-accuracy surrogate for the smooth background map $\mathcal{F}_{\mathrm{bg}}$. Its output can be interpreted as globally coupled feature representations that serve as tokens for a subsequent transformer. From a physical perspective, this spectral token mixing encodes background wave propagation through global spatial coupling. The resulting feature maps are then fed to a vision transformer, which combines these globally mixed representations to capture the complex scattering produced by the obstacle (e.g. salt body).
This idea is similar to Adaptive Fourier Neural Operators (AFNOs) \cite{Guibas2021AdaptiveFN}, which employ FNO-style spectral mixing as a token mixer for vision transformers in non-operator-learning settings.

\begin{figure*}[t]
    \centering
    \includegraphics[width=\linewidth]{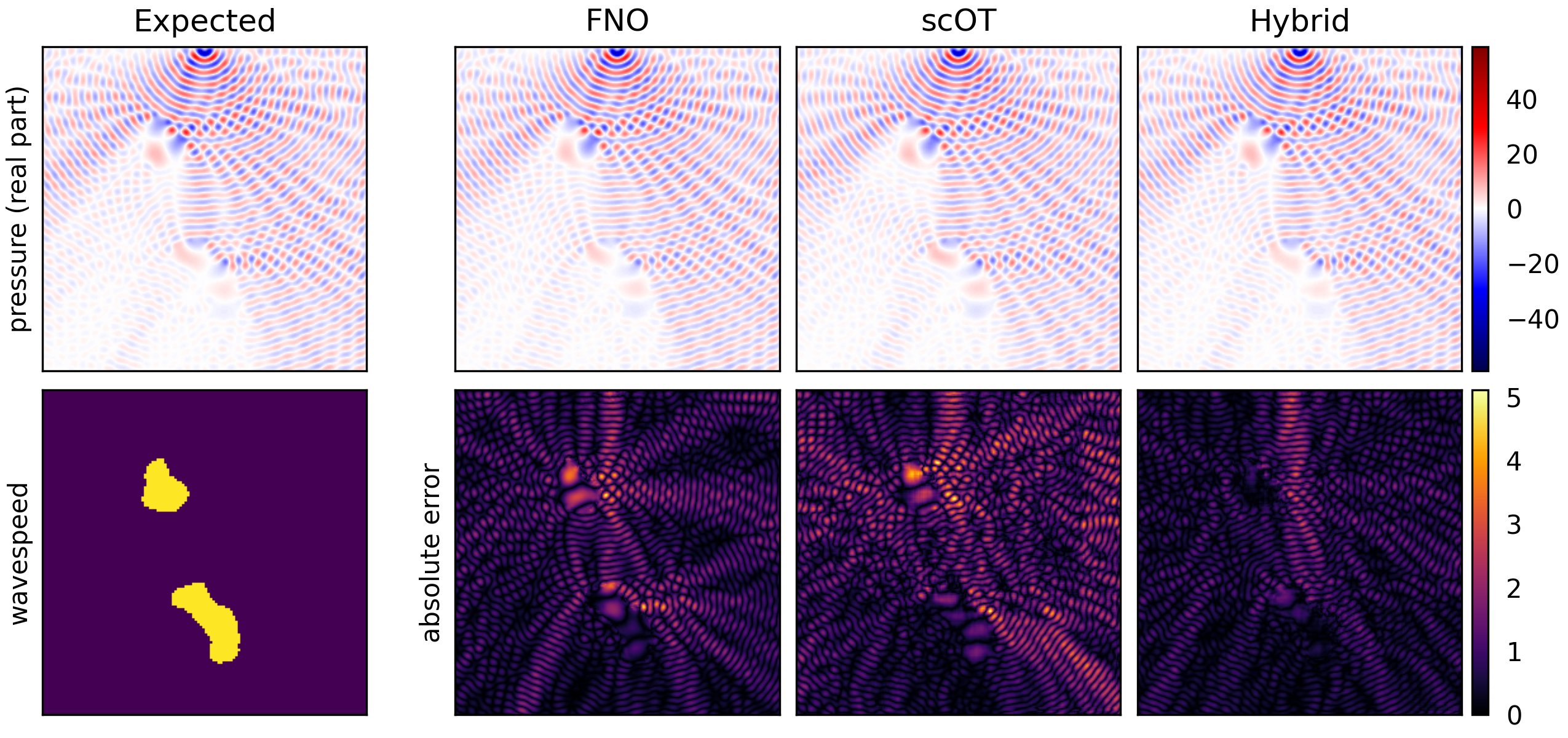}
\caption{Helmholtz wavefield predictions using different architectures. (Top row) Real part of the predicted pressure field $p(\mathbf{x}, \omega)$ for the expected (numerical) solution and three architectures: FNO, scOT, and (\emph{ours}) Hybrid. (Bottom row) Absolute prediction errors relative to the expected solution. The velocity model (bottom left) contains two obstacles (salt bodies) in a homogeneous background. The Hybrid architecture demonstrates superior accuracy with significantly reduced error compared to FNO and scOT approaches, particularly in capturing fine-scale wavefield features around the obstacles. Additional result visualizations are presented in appendix \ref{appendix:results}.}
\label{fig:wavefield_predictions}
\end{figure*}

\begin{figure*}[t]
    \centering
    \includegraphics[width=\linewidth]{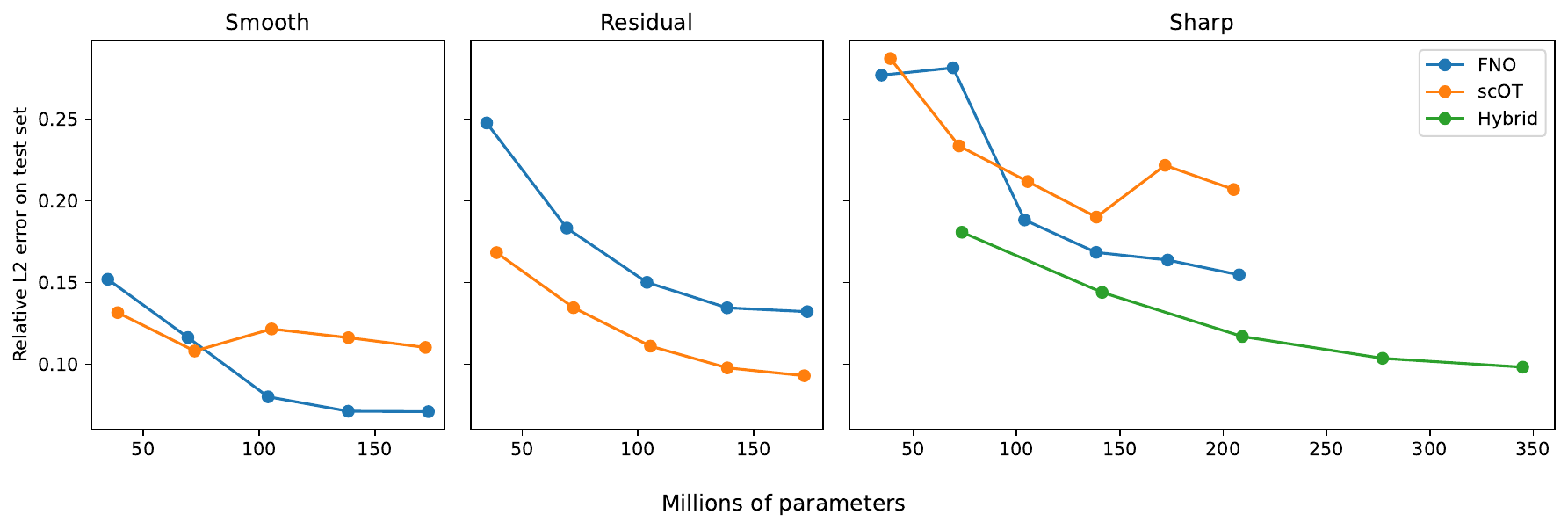}
    \caption{Performance of FNO, scOT, and Hybrid architectures with different parameter sizes across three different learning tasks: (L to R) Smooth ($v_{\mathrm{bg}} \mapsto p_{\mathrm{bg}}$), Residual data ($p_{\mathrm{bg}}(\cdot, \omega), \delta v \mapsto \delta p$), and full end-to-end Helmholtz solution on sharp wavespeeds ($v \mapsto p$). The Hybrid architecture (this work) enables end-to-end learning of the forward operator in the high-contrast scattering case.}
    \label{fig:parameter_performance}
\end{figure*}

\subsection{Transformer design for the scattering}

We propose that vision transformers are a natural surrogate model for learning the corrector $\mathcal{F}_{\mathrm{sc}}$ since the attention mechanism is well-suited for the strongly spatially dependent structure of the scattered field.  
Although vision transformers have appeared in operator-learning architectures, there is limited justification for interpreting patch-based tokenization as a consistent discretization of an underlying continuous operator.  We therefore outline a framework that provides such an interpretation for patch-based vision transformers.

A transformer operates on a sequence of tokens (i.e. a context) and updates this context through a sequence-to-sequence map.
In order to cast this into a functional framework suitable for operator learning, the sequences of tokens must arise as discretizations of an underlying function on a spatial domain. 
A standard transformer acts on a sequence of tokens, but for operator learning these tokens are viewed as discretizations of an underlying function on a spatial domain. Refining the tokenization corresponds to refining the discretization, with the continuous function recovered in the limit of infinitely many tokens. This viewpoint is formalized by the measure-theoretic transformer framework \cite{deHoop2024universal-in-context}, which interprets classical transformers operating on discrete token sequences as discretizations of operators acting on measure-valued contexts.

From this perspective, a function $f(x)$ can be identified with its graph $(x,f(x))$ endowed with a canonical pushforward measure. Patch-based tokenizations with positional encodings correspond to discrete samples of this graph measure, while self-attention induces interactions between these samples that approximate an operator acting on the underlying continuous representation. In our setting, $f(x)$ represents the background wavefield $p_{\mathrm{bg}}(\bx,\omega)$, and the resulting attention mechanism captures geometry-dependent scattering interactions induced by high-contrast obstacles. The patching operation itself can be viewed as a local projection that converges to the identity as the patch size tends to zero, so that the patch-based vision transformer recovers a consistent discretization of a continuum scattering operator acting on the wavefield.

Guided by this framework, we approximate the corrector using a patch-based vision transformer architecture. 
The patching operator transforms the input function into a piecewise constant function, which is constant within patches (subdivisions of the domain $\Omega$), by taking weighted averages and then transforming these piecewise constant values into a $C$-dimensional latent space resulting in output $v \in C(\Omega; \mathbb{R}^C)$. 


In order to mitigate the complexity associated with attention, we use the Swin-style transformer. Swin V2 \cite{Swin-v2} introduces windowed self-attention to make transformers scalable, hierarchical, and inductively biased toward locality, while still retaining global modeling via shifting and multi-scale structure. Shifting creates cross-window connectivity across layers, allowing information to propagate globally through a sequence of local attention operations. One can think of this as controlled information percolation.
Windowing is introduced to make self-attention computationally feasible, spatially local, and hierarchically compositional for images (one can loosely think of this as Schwarz-type domain decomposition in disguise). Self-attention is computed only inside each window so that the complexity becomes linear in image (i.e. domain) size for a fixed window size.


%% file: 05_Experiments.tex
\section{Experiments}

To evaluate our proposed hybrid architeture, we trained a sequence of FNOs and scOT transformers of varying sizes on three different learning tasks:
\begin{itemize}
    \item \textbf{Smooth} task: Approximate the Helmholtz solution operator on smooth wavespeeds: $\mathcal{F}_\text{bg}(s, v_\text{bg}) = p_\text{bg}$.
    \item \textbf{Residual} task: Approximate the scattering corrector $\mathcal{F}_\text{sc}(p_\text{bg}, \delta v) = \delta p$.
    \item \textbf{Sharp} task: Approximate the full Helmholtz solution operator $\mathcal{F}(s, v) = p$.
\end{itemize}

We validate empirically that FNOs consistently outperform scOT transformers on the smooth task; conversely, scOT transformers consistently outperform FNOs on the residual task. This demonstrates that FNO and scOT are correctly suited to their subtasks in the hybrid architecture presented in Section \ref{sec:architecture}. Furthermore, we observe that our hybrid models are consistently more accurate than either FNO or scOT networks of comparable size on the end-to-end sharp task.
Details of the training protocol are presented in Appendix~\ref{appendix:training}.




\subsection{Accuracy}

Figure \ref{fig:wavefield_predictions} presents qualitative comparisons of Helmholtz wavefield predictions across the three architectures on a high-contrast test sample containing two salt bodies. The top row shows the real part of the predicted pressure field $p(\mathbf{x}, \omega)$ for the expected numerical solution and the three learned architectures: FNO, scOT, and Hybrid. All three models capture the overall wave propagation pattern, including the interference fringes and phase structure characteristic of the frequency regime. However, the bottom row, which displays absolute prediction errors relative to the expected solution, reveals significant differences in accuracy. Both the FNO and scOT exhibit substantial errors, particularly in regions surrounding the salt bodies where strong scattering, reflections, and diffractions occur. The FNO's convolution-based design struggles with the long-range spatial dependencies induced by scattering, while scOT performs slightly worse despite using attention mechanisms, suggesting that direct end-to-end prediction of the complete wavefield remains challenging even with global receptive fields. On the other hand, the Hybrid architecture achieves significantly lower errors across the domain, with substantially reduced error around the salt bodies and improved capture of fine-scale wavefield features in the scattered field.  

\subsection{Parameter efficiency}

Figure \ref{fig:parameter_performance} demonstrates the parameter efficiency of FNO and scOT on the smooth (left) and residual (center) tasks, as well as FNO, scOT, and our hybrid architecture on the high-contrast, sharp task (right). For each task, we trained 5 FNOs with 2, 4, 6, 8, and 10 layers, 64 hidden features, and 64 modes, using the official PyTorch neuraloperator library \cite{kossaifi2025librarylearningneuraloperators}. Likewise, we trained 5 scOT transformers with 2, 4, 6, 8, and 10 transformer blocks in each attention head (``depths''), an embedding dimension of 90, and all other parameters identical to the Poseidon-B architecture \cite{Herde2024Poseidon}. These configurations lead to comparable parameter counts between the two architectures.

For the smooth task, both architectures learn the map from smoothed wavespeed to background pressure $(v_{\mathrm{bg}} \mapsto p_{\mathrm{bg}})$, where minimal discontinuities lead to comparable performance across FNO and scOT. The FNO shows a distinct advantage at higher parameter counts due to its inductive bias for smooth propagation. In the residual regime (center panel), FNO and scOT learn the map $(p_{\mathrm{bg}}(\cdot,\omega), \delta v \mapsto \delta p)$, predicting the pressure residual from the background field and wavespeed residual. As scattering intensity increases, scOT begins to outperform the FNO, achieving lower relative $L^2$ error across parameter budgets. The hybrid architecture, composed of an FNO trained on the smooth task and a scOT transformer trained on the residual task, outperforms either FNO or scOT to recover the full end-to-end map $(v \mapsto p)$ in the presence of high-contrast salt bodies that induce strong reflections, refractions, and multiple scattering events. Our hybrid architecture constructed in this way outperforms both FNO and scOT models trained on the sharp task, even though the hybrid model never sees paired data $(v, p)$.
Additionally, the hybrid exhibits favorable scaling behavior: as model capacity increases, the loss decreases sharply relative to both baselines. 
This quantitative assessment confirms the qualitative results from Figure \ref{fig:wavefield_predictions}: the decomposition into smooth background propagation (FNO) and high-contrast scattering correction (transformer) enables the Hybrid model to more accurately resolve the complex wave phenomena in strongly scattering, high-contrast regimes.

%% file: 06_Discussion.tex
\section{Discussion and Future Work}






The experimental results validate our central hypothesis that the high-contrast wave scattering constitutes a distinct regime where decomposition-based hybrid architectures outperform FNO and scOT.
Both architectures struggle on the high-contrast task when learning the full end-to-end map directly, with transformers performing slightly worse than FNOs despite their global receptive fields. This suggests that neither architecture alone can efficiently resolve the disparate spatial scales and regularity properties present when background propagation and scattering must be learned simultaneously.  In our hybrid model, FNOs and scOT exhibit complementary strengths: FNOs effectively capture smooth background propagation, while transformers excel at modeling the scattering correction $\mathcal{F}_{\mathrm{sc}}$ when the background field is provided as input. The superior performance of scOT on the high-contrast corrector indicates that, within the hybrid architecture, the primary role of the FNO is to generate a globally coupled feature representation that is subsequently refined by attention. 
This validates our general design principle for high-contrast wave scattering: global spectral coupling followed by localized, content-adaptive refinement.

Several directions emerge from this work. 
Extending quantitative out-of-distribution generalization estimates to transformer architectures could inform principled regularization strategies and provide insight into when transformers outperform classical neural operators in high-contrast regimes. Architectural refinements such as stochastic depth and curriculum learning strategies may improve robustness and performance. The hybrid framework's accurate prediction of wavefields and derivatives also enables direct application to inverse problems via adjoint-state methods, where access to Fr\'echet derivatives is essential for gradient-based inversion algorithms.

%% file: 07_appendix.tex
\appendix
\onecolumn

\section{Lippmann–Schwinger Formulation for the Helmholtz Operator}
\label{app:greens-function}

We briefly recall the definition of the Green's function $G_s$ associated with
the background velocity $v_s(\boldsymbol{x})$ and the Helmholtz equation~\eqref{eq:helmholtz}.

For a fixed angular frequency $\omega$, the background Helmholtz operator is
\begin{equation}
    \mathcal{L}_s(\omega) p
    \;\coloneqq\;
    \left[ \Delta + \frac{\omega^2}{v_s(\boldsymbol{x})^2} \right] p(\boldsymbol{x},\omega),
\end{equation}
acting on functions $p:\Omega \to \mathbb{C}$ satisfying the same boundary
conditions as in Secction~\ref{sec:helmholtz}. The \emph{background Green's
function} $G_s(\boldsymbol{x},\boldsymbol{y},\omega)$ is defined as the
distributional solution of
\begin{equation}
    \mathcal{L}_s(\omega)\, G_s(\boldsymbol{x},\boldsymbol{y},\omega)
    = -\delta(\boldsymbol{x}-\boldsymbol{y}),
    \qquad \boldsymbol{x} \in \Omega,
\end{equation}
with $\boldsymbol{x}$ satisfying the same boundary conditions on $\partial\Omega$
as $p$.

Given a source term $s(\boldsymbol{y},\omega)$, the corresponding background
field $p_s(\boldsymbol{x},\omega)$ solves
\[
    \mathcal{L}_s(\omega) p_s(\boldsymbol{x},\omega)
    = -s(\boldsymbol{x},\omega),
\]
and admits the integral representation
\begin{equation}
    p_s(\boldsymbol{x},\omega)
    =
    \int_\Omega G_s(\boldsymbol{x},\boldsymbol{y},\omega)\,
    s(\boldsymbol{y},\omega)\, \mathrm{d}\boldsymbol{y}.
\end{equation}

For the full velocity $v(\boldsymbol{x})$ we write
\[
    v^{-2}(\boldsymbol{x})
    = v_s^{-2}(\boldsymbol{x}) + \delta m(\boldsymbol{x}),
    \qquad
    \delta m(\boldsymbol{x}) = v^{-2}(\boldsymbol{x}) - v_s^{-2}(\boldsymbol{x}),
\]
and let $p(\boldsymbol{x},\omega)$ denote the corresponding total field solving
equation~\eqref{eq:helmholtz}. Standard Green's function identities then yield the
Lippmann Schwinger equation
\begin{equation}
    p(\boldsymbol{x},\omega)
    =
    p_s(\boldsymbol{x},\omega)
    +
    \int_\Omega
        G_s(\boldsymbol{x},\boldsymbol{y},\omega)\,
        \delta m(\boldsymbol{y})\, p(\boldsymbol{y},\omega)\,
    \mathrm{d}\boldsymbol{y}.
\end{equation}

\section{Models for training data: from GRFs to obstacles}
\label{app:salt_image_blur}

We describe how we construct the sharp velocity model $v(\boldsymbol{x})$
and the smoothed background model $v_s(\boldsymbol{x})$.

\paragraph{Binary Salt Mask from a 3D GRF.}
We first generate a three-dimensional Gaussian random field
$\phi(z,y,x)$ on a regular grid of size
$N_z \times N_y \times N_x = 256 \times 256 \times 256$ using a spectral Matérn-type covariance
with correlation lengths $(\ell_x,\ell_y,\ell_z) = (40,40,40)$ and smoothness
parameter $\nu_{\text{GRF}} =1.6$. 

We then select a single 2D slice
$\chi_{\mathrm{salt}}^{2\mathrm{D}}(y,x) \in \{0,1\}^{N_y\times N_x}$
from $\chi_{\mathrm{salt}}^{3\mathrm{D}}$ with sufficiently large salt
area and possibly multiple disjoint blobs. This 2D mask defines the
salt geometry on an image grid with physical coordinates
\[
    x \in [0,L_x], \qquad y \in [0,L_y].
\]

\paragraph{Background and Sharp Velocity.}
On the Helmholtz solver grid of size $n_y \times n_x$ with
coordinates
\[
    x_i = \frac{i}{n_x-1}L_x, \quad i=0,\dots,n_x-1,
    \qquad
    y_j = \frac{j}{n_y-1}L_y, \quad j=0,\dots,n_y-1,
\]
we define a constant background velocity $v_{bg}(x,y)$ and a constant
salt velocity $v_{\mathrm{salt}}$.
The 2D mask $\chi_{\mathrm{salt}}^{2\mathrm{D}}$ is sampled onto this
solver grid by nearest-neighbor interpolation, yielding
$\chi_{\mathrm{salt}}(x,y) \in \{0,1\}$.

The sharp (high-contrast) velocity model is then
\begin{equation}
    v(x,y)
    \;=\;
    v_{\mathrm{salt}}\,\chi_{\mathrm{salt}}(x,y)
    \;+\;
    v_{\mathrm{bg}}(x,y)\,\big(1 - \chi_{\mathrm{salt}}(x,y)\big).
    \label{eq:sharp-velocity}
\end{equation}

\paragraph{Gaussian Blurring of the Salt Interface.}
We smooth the velocity by convolving
$\chi_{\mathrm{salt}}^{2\mathrm{D}}$ with an anisotropic Gaussian
kernel in the \emph{image} grid. Let
$\Delta x_{\mathrm{img}}$ and $\Delta y_{\mathrm{img}}$ denote the
image-grid spacings in the $x$- and $y$-directions, and let
$\sigma_{\mathrm{salt}}$ (in meters) be the desired physical smoothing
scale along the interface. The corresponding pixel-standard deviations are
\[
    \sigma_x^{\mathrm{pix}} = \frac{\sigma_{\mathrm{salt}}}{\Delta x_{\mathrm{img}}},
    \qquad
    \sigma_y^{\mathrm{pix}} = \frac{\sigma_{\mathrm{salt}}}{\Delta y_{\mathrm{img}}}.
\]
We then form a smoothed salt ``fraction'' field
\begin{equation}
    \alpha_{\mathrm{img}}(y,x)
    \;=\;
    \big(G_{\sigma_y^{\mathrm{pix}},\sigma_x^{\mathrm{pix}}}
        * \chi_{\mathrm{salt}}^{2\mathrm{D}}\big)(y,x),
\end{equation}
where $G_{\sigma_y^{\mathrm{pix}},\sigma_x^{\mathrm{pix}}}$ is a
separable Gaussian kernel and $*$ denotes convolution on the image grid.
We clip $\alpha_{\mathrm{img}}$ to $[0,1]$.
This field is then interpolated to the solver grid,
yielding $\alpha(x,y) \in [0,1]$.

The smoothed background velocity model is then
\begin{equation}
    v_s(x,y)
    \;=\;
    v_{\mathrm{salt}}\,\alpha(x,y)
    \;+\;
    v_{\mathrm{bg}}(x,y)\,\big(1 - \alpha(x,y)\big).
    \label{eq:smoothed-velocity}
\end{equation}

\subsection{Matérn Gaussian Random Fields}
\label{app:Matern}

To generate diverse yet controllable subsurface geometries, we model
salt bodies as level sets of a zero–mean Gaussian Random Field (GRF)
$\phi(x)$ defined on the computational domain
$\Omega \subset \mathbb{R}^2$.
The field is specified by a Matérn covariance kernel
\cite{rasmussen2006gp}, which provides separate control over variance,
correlation length, and smoothness.

Let $x,x'\in\Omega$. The Matérn covariance between
$\phi(x)$ and $\phi(x')$ is
\begin{equation}
    \mathrm{Cov}\!\left[\phi(x),\phi(x')\right]
    \;=\;
    k_{\text{Matérn}}(x,x')
    \;=\;
    \sigma^2 \,\frac{2^{1-\nu}}{\Gamma(\nu)}
    \left( \frac{\|x-x'\|}{\ell} \right)^{\nu}
    K_\nu\!\left( \frac{\|x-x'\|}{\ell} \right),
    \label{eq:app-matern}
\end{equation}
where $\sigma^2$ is the marginal variance, $\ell>0$ is the correlation
length, $\nu>0$ is the smoothness parameter, $\Gamma$ is the Gamma
function, and $K_\nu$ is the modified Bessel function of the second
kind. Larger values of $\ell$ increase the typical size of features in
$\phi$, while larger $\nu$ produce smoother level sets.

We discretize $\Omega$ on an $H\times W$ Cartesian grid
$\{x_{ij}\}_{i,j=1}^{H,W}$, stack all grid points into a vector
$\mathbf{x} \in \mathbb{R}^{HW}$, and define the covariance matrix
$K \in \mathbb{R}^{HW \times HW}$ with entries
\[
    K_{pq}
    \;=\;
    k_{\text{Matérn}}(x_p,x_q),
\]
where $p,q$ index grid points in a fixed ordering.
A GRF realization is then obtained by sampling
\[
    \boldsymbol{\phi} \sim \mathcal{N}(0, K),
\]
and reshaping $\boldsymbol{\phi}\in\mathbb{R}^{HW}$ back to the
$H\times W$ grid.
In practice, we fix $\sigma^2 = 1$ and choose $(\ell,\nu)$ so that the
resulting level sets produce blob-like inclusions with typical diameter
comparable to several wavelengths at the frequencies of interest.

\subsection{Salt Bodies from Thresholded GRFs}
\label{app:salt_bodies}

Each realization $\phi(x)$ defines a random salt geometry via
thresholding.
Given a target salt volume fraction $\rho_{\text{salt}} \in (0,1)$, we
choose a threshold $\tau$ such that
\[
    \frac{1}{|\Omega|}
    \big|\{x \in \Omega : \phi(x) > \tau\}\big|
    \approx
    \rho_{\text{salt}}.
\]
We then define the binary salt indicator
\[
    \chi_{\text{salt}}(x)
    \;=\;
    \mathbf{1}\{\phi(x) > \tau\},
\]
so that $\chi_{\text{salt}}(x)=1$ inside salt and $0$ in the
background.
On the discrete grid, this amounts to thresholding the sampled field
$\boldsymbol{\phi}\in\mathbb{R}^{H\times W}$ elementwise.

The individual \emph{salt bodies} are the connected components of the
set $\{x \in \Omega : \chi_{\text{salt}}(x)=1\}$.
For numerical robustness and to avoid unrealistically small features,
we optionally remove components whose area falls below a prescribed
minimum and fill small gaps in thin layers using standard
morphological operations (opening/closing) on the indicator image.

Given the salt indicator, we construct a piecewise-constant velocity
model
\begin{equation}
    v(x)
    \;=\;
    v_{\text{bg}}
    + \bigl(v_{\text{salt}} - v_{\text{bg}}\bigr)\,
      \chi_{\text{salt}}(x),
    \label{eq:app-salt-velocity}
\end{equation}
where $v_{\text{salt}}$ is the velocity assigned to salt and
$v_{\text{bg}}$ the background velocity.
In all experiments, we use
\[
    v_{\text{salt}} = 4.5~\mathrm{km/s},
    \qquad
    v_{\text{bg}}   = 1.5~\mathrm{km/s}.
\]
Thus each connected salt body has constant high velocity
$v_{\text{salt}}$, embedded in a homogeneous background of velocity
$v_{\text{bg}}$.

\section{Training protocols}
\label{appendix:training}

\subsection{Data splits}
We generated 50,000 high-contrast wavespeed images ($v$) of size 256x256 using the procedure in appendix \ref{app:salt_image_blur}. The ground-truth pressure fields were computed using the numerical solver hawen \cite{Faucher2021} using the boundary conditions from Section \ref{sec:helmholtz}, a point source located at the center of the free boundary and 40Hz frequency. We split this dataset into 40,000 training pairs, 5,000 validation pairs, and 5,000 test pairs. The validation set was used to select the optimal models from training runs, and the test set was used to evaluate models.

\subsection{Training procedure}
All models were trained with the AdamW optimizer \cite{loshchilov2019decoupledweightdecayregularization} for 100 epochs using a cosine learning rate decay scheduler \cite{loshchilov2017sgdrstochasticgradientdescent} with 5 epoch linear warm-up period and a max learning rate of $10^{-3}$.

\subsection{Loss}
All models were trained to optimize the relative $L_2$ loss:
Let $P_{ij}$ and $\widehat{P}_{ij}$ denote the true and
predicted complex values at grid point $(i,j)$.
The relative $L_2$ error (Rel-$L_2$) is defined as
\begin{equation}
  \mathrm{Rel}\text{-}L_2
  = \frac{\|\widehat{P} - P\|_2}{\|P\|_2},
\end{equation} 


\clearpage
\section{Experiment Data Visualization}

\begin{figure}[h!]
    \centering
        \includegraphics[width=0.75\linewidth,trim={1.5cm 2cm 1.5cm 2.2cm},clip]{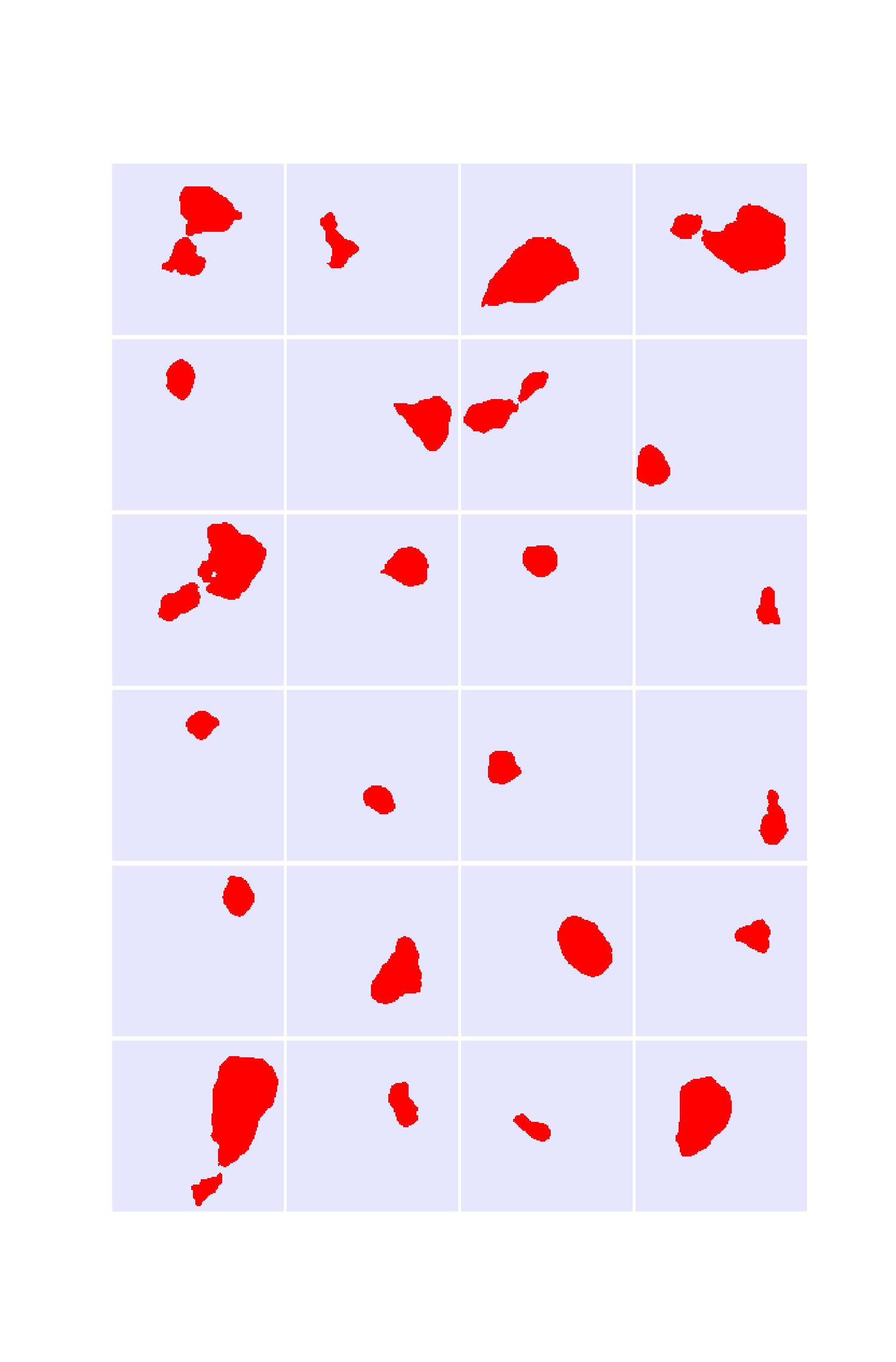}
    \caption{Input wavespeeds sampled randomly from the training data. The red obstacle regions (e.g. salt) have a wavespeed of 4.5~km/s, and the blue background regions have a wavespeed of 1.5~km/s.}
    \label{fig:example-velocity-fields}
\end{figure}

\begin{figure}[h!]
    \centering
    \includegraphics[width=0.75\linewidth,trim={1.5cm 2cm 1.5cm 2.2cm},clip]{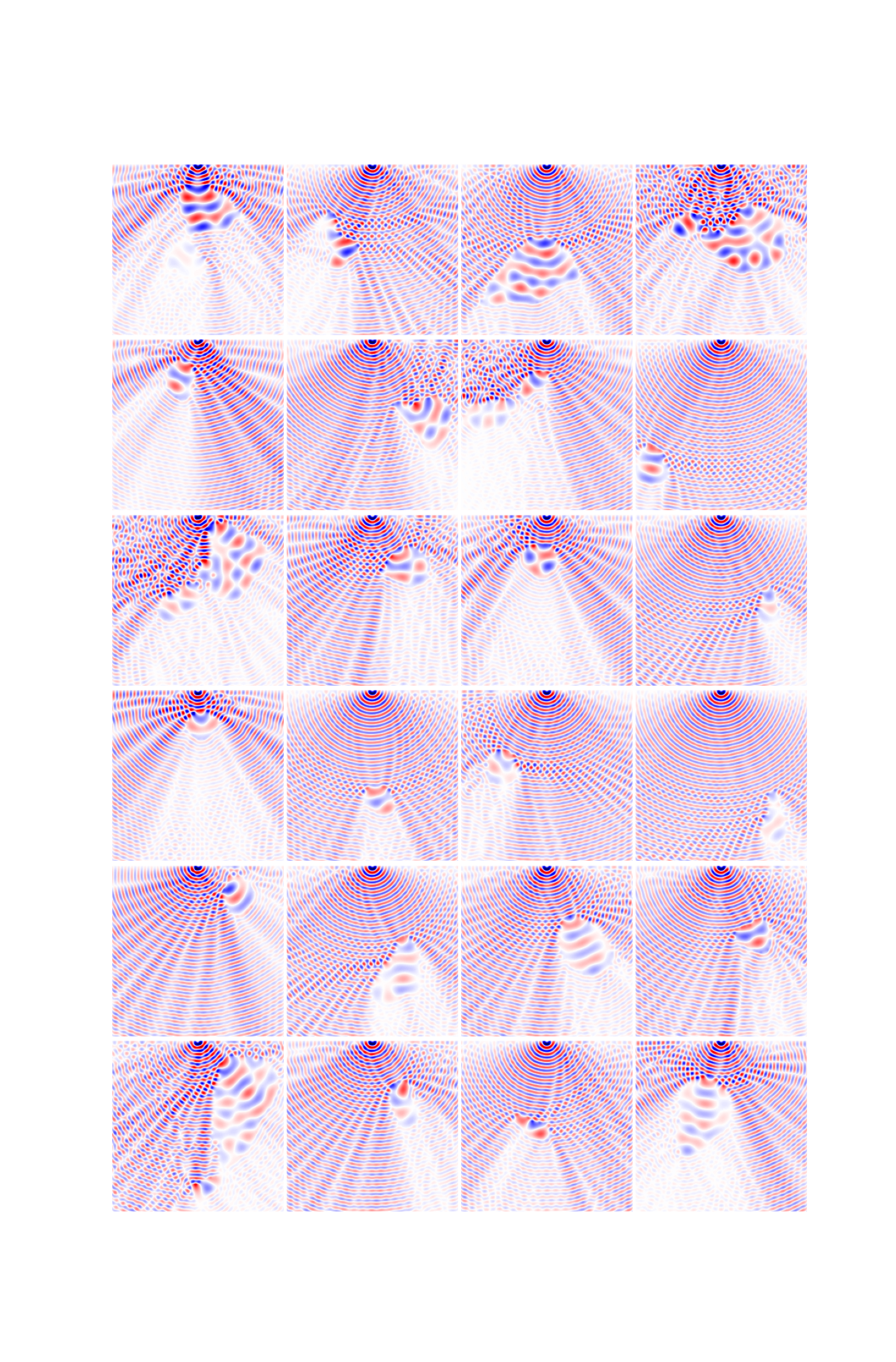}
    \caption{Real parts of the target pressure fields corresponding to the wavespeeds from Figure \ref{fig:example-velocity-fields}.}
\end{figure}

\clearpage
\section{Additional result visualizations}
\label{appendix:results}

This appendix contains additional result visualizations on samples randomly selected from the test dataset.

\begin{figure}[h!]
    \centering
    \includegraphics[width=0.8\linewidth]{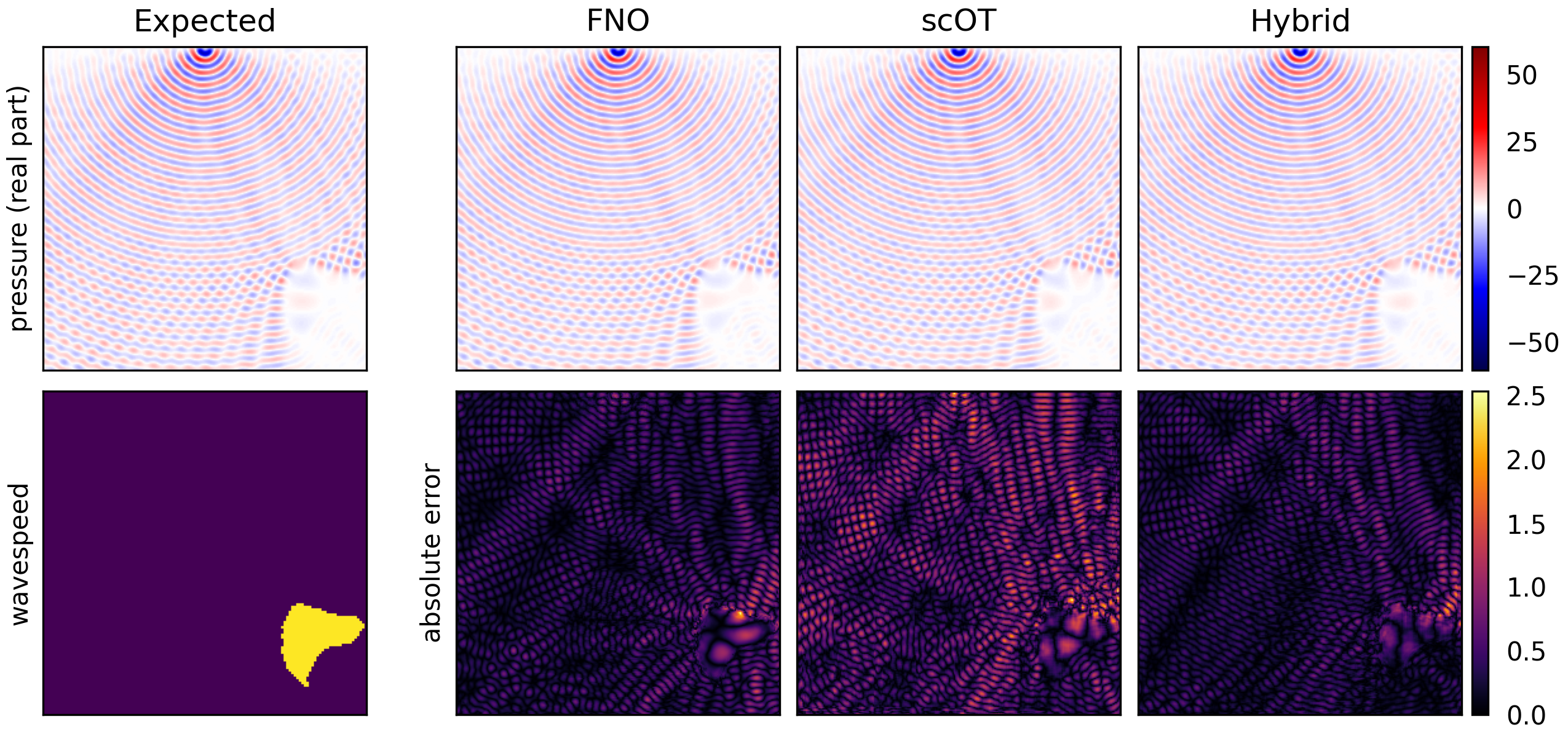}
\end{figure}

\begin{figure}[h!]
    \centering
    \includegraphics[width=0.8\linewidth]{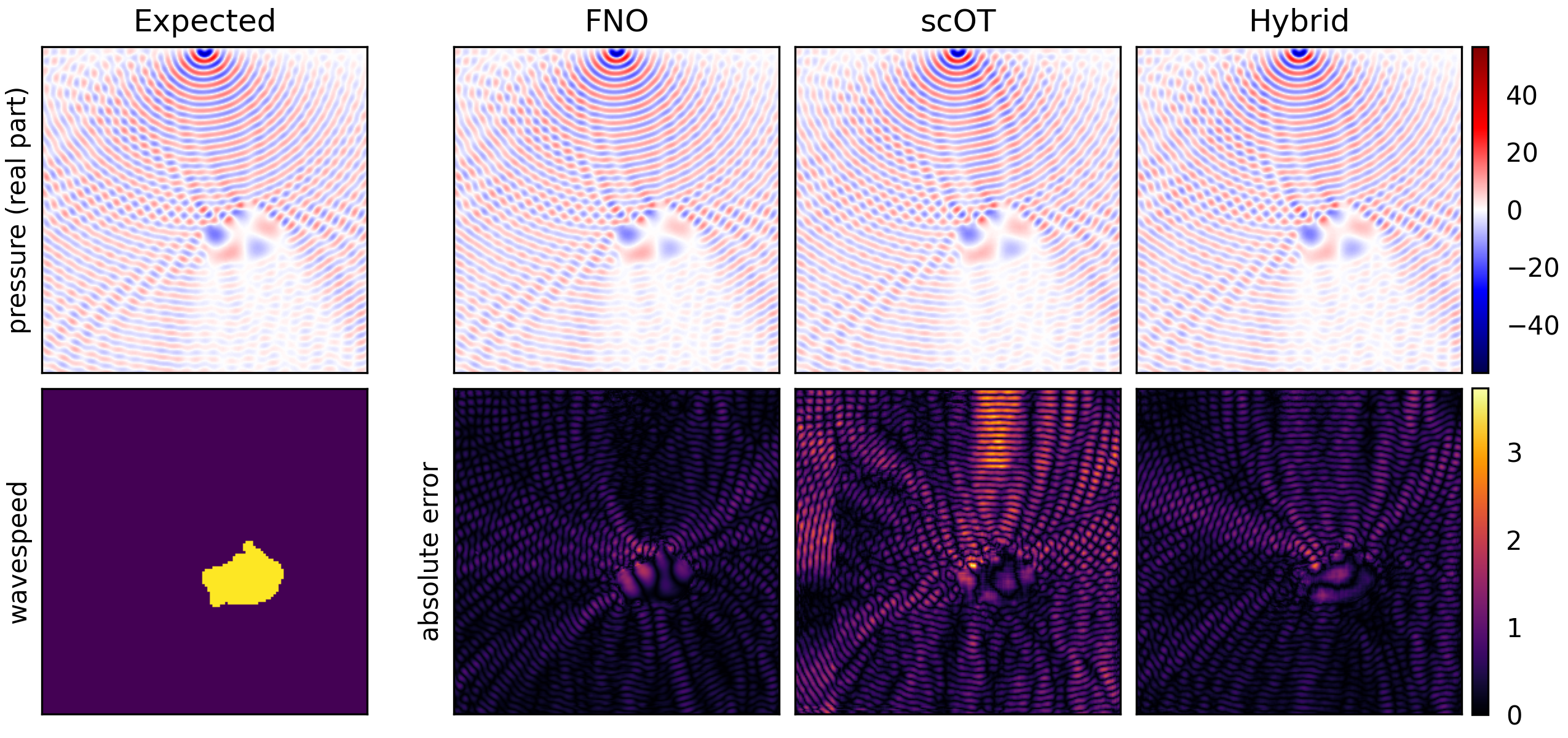}
\end{figure}

\begin{figure}[h!]
    \centering
    \includegraphics[width=0.8\linewidth]{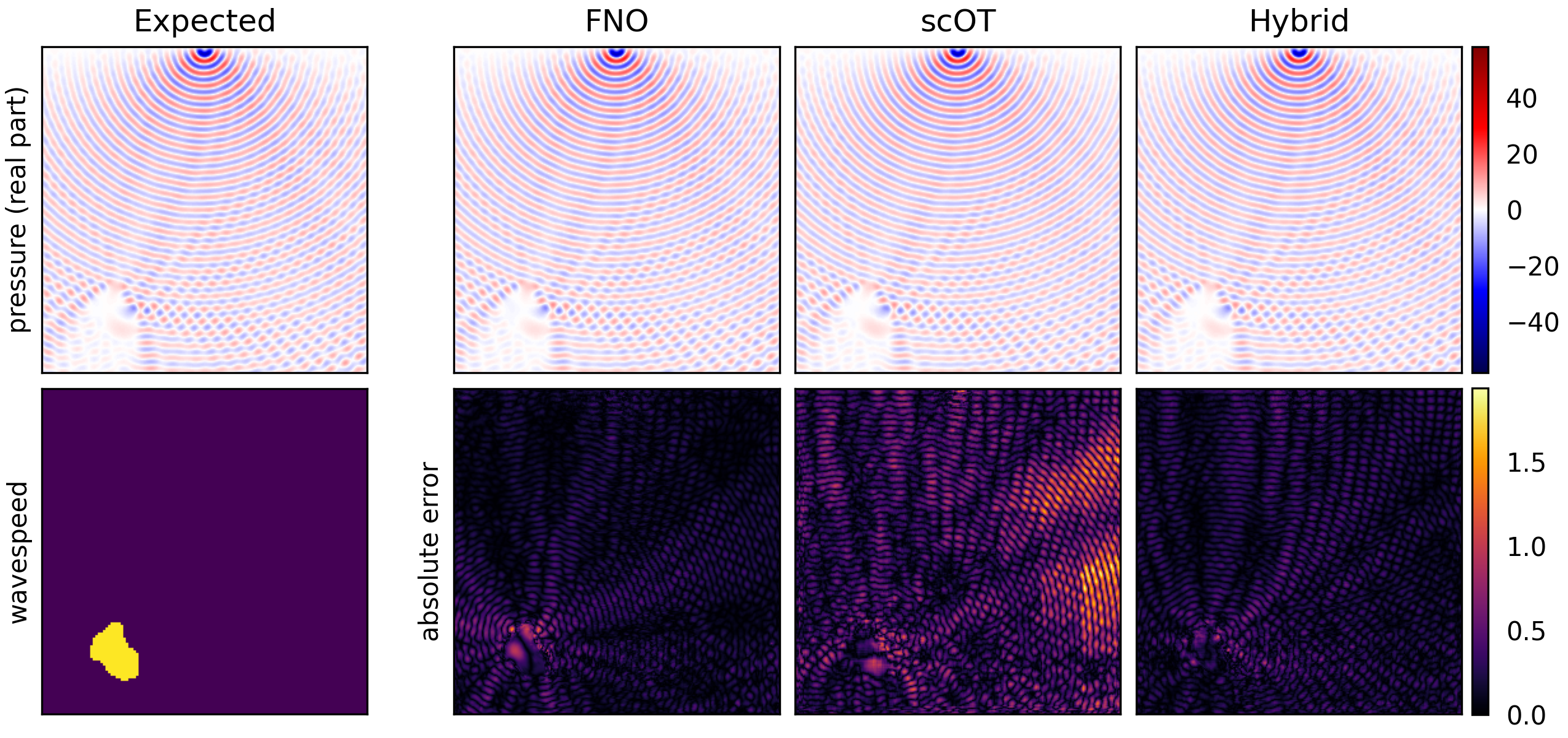}
\end{figure}

\begin{figure}[h!]
    \centering
    \includegraphics[width=0.8\linewidth]{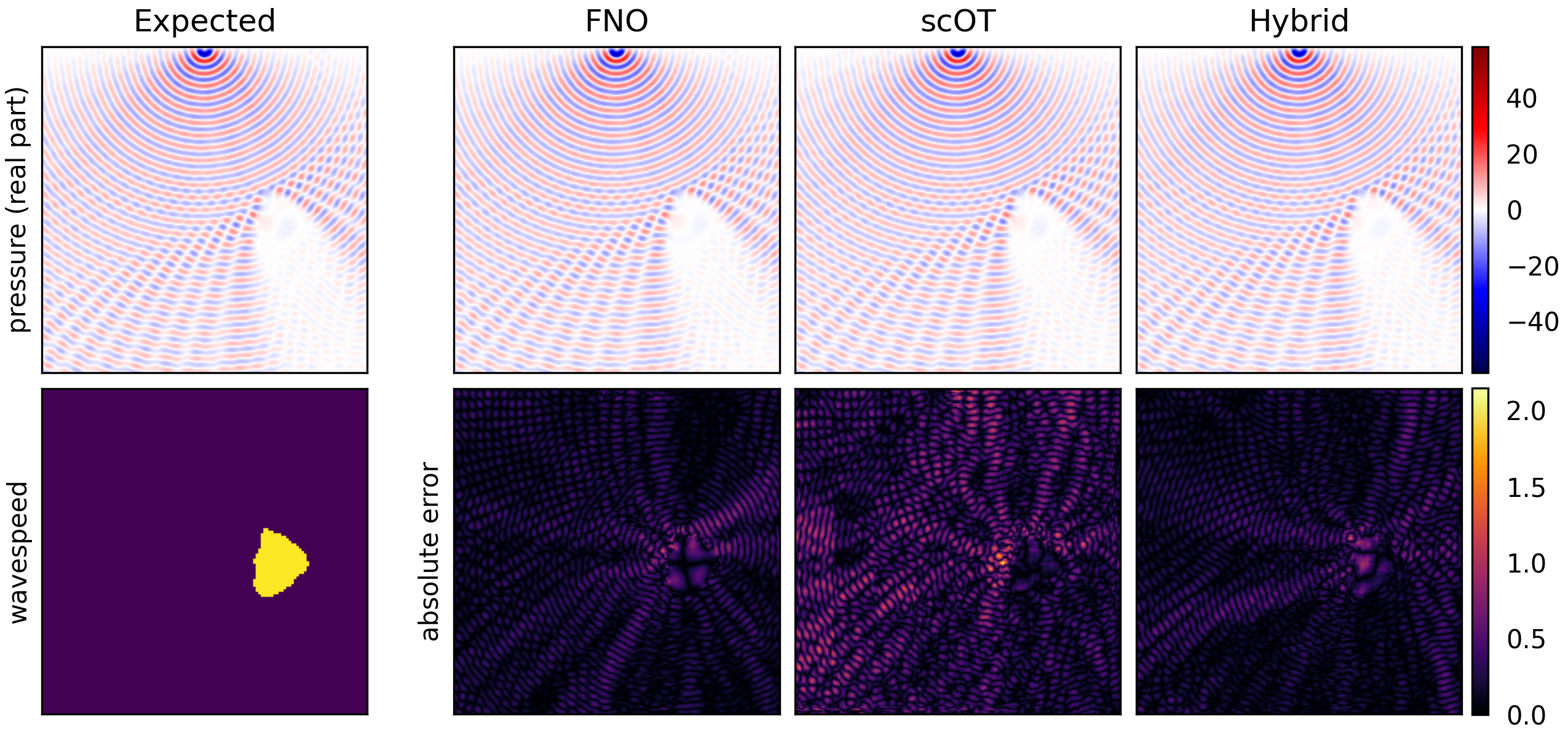}
\end{figure}

\begin{figure}[h!]
    \centering
    \includegraphics[width=0.8\linewidth]{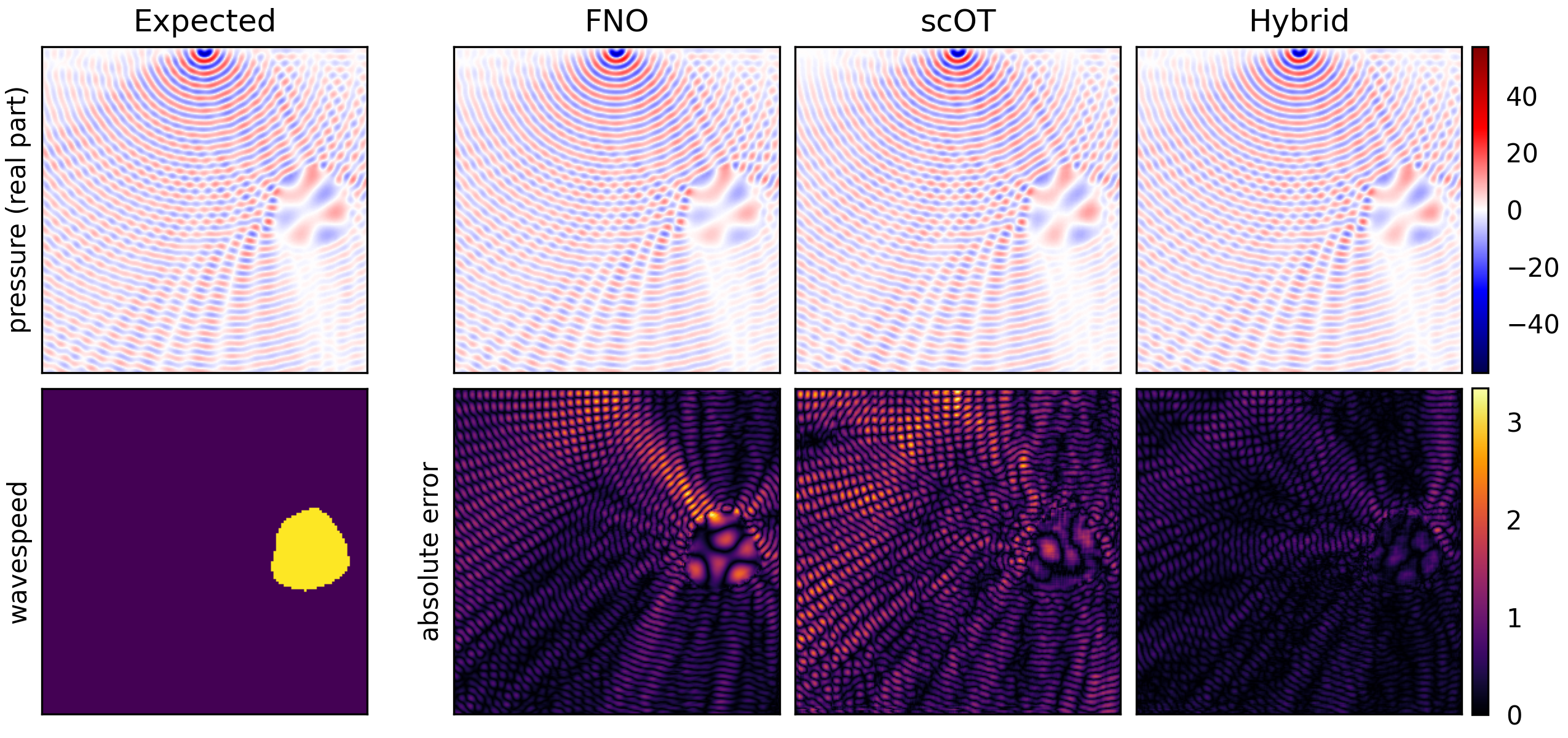}
\end{figure}

\begin{figure}[h!]
    \centering
    \includegraphics[width=0.8\linewidth]{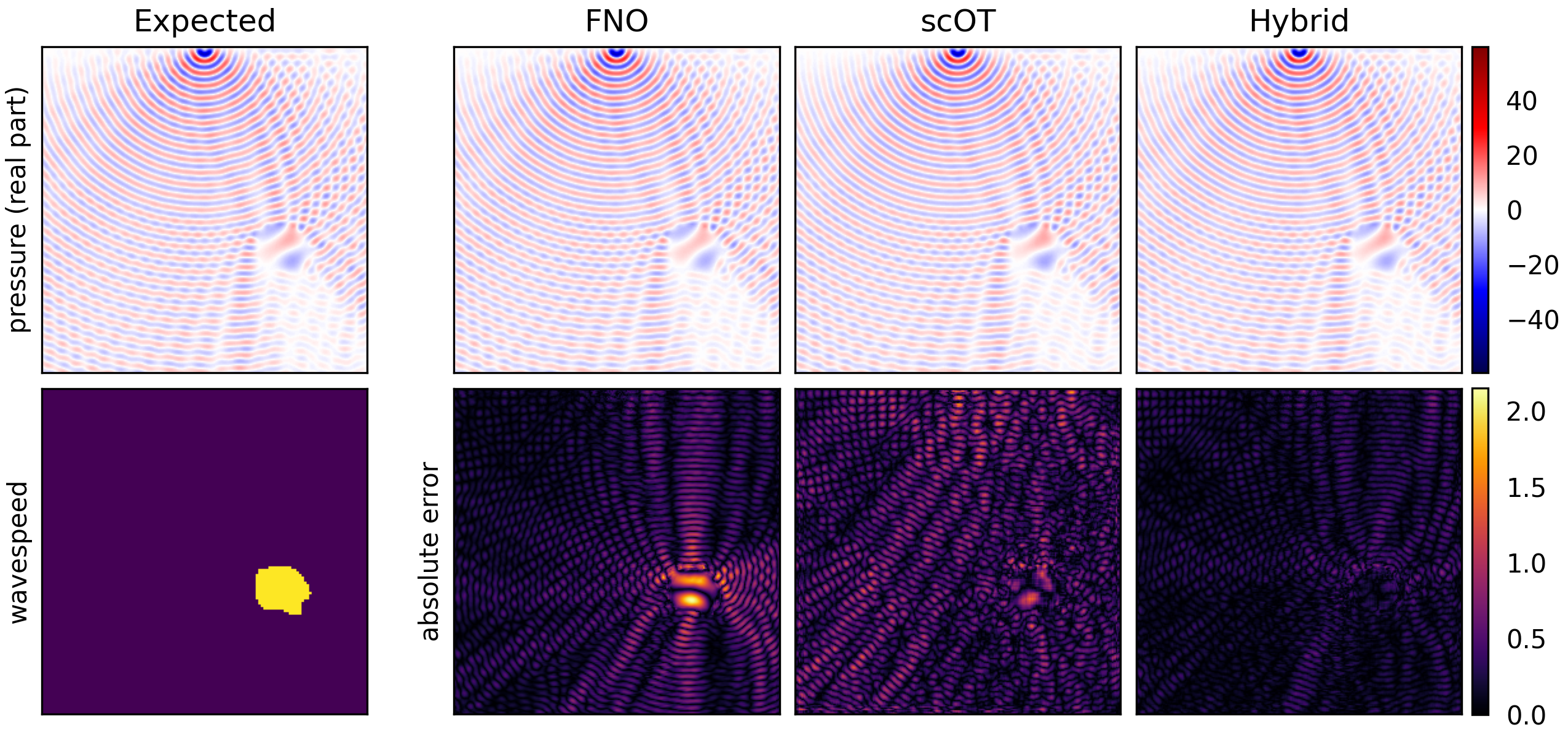}
\end{figure}

\begin{figure}[h!]
    \centering
    \includegraphics[width=0.8\linewidth]{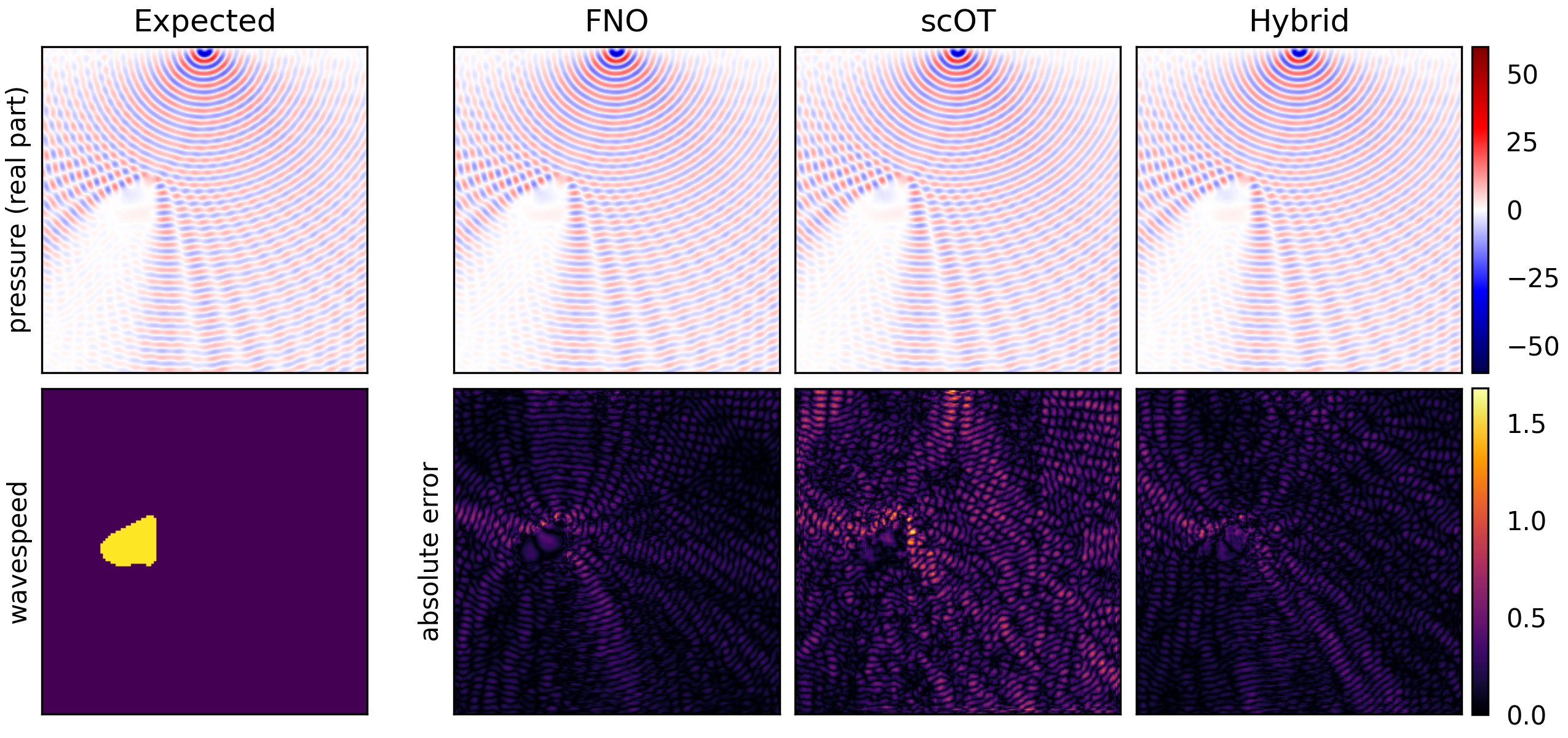}
\end{figure}

\begin{figure}[h!]
    \centering
    \includegraphics[width=0.8\linewidth]{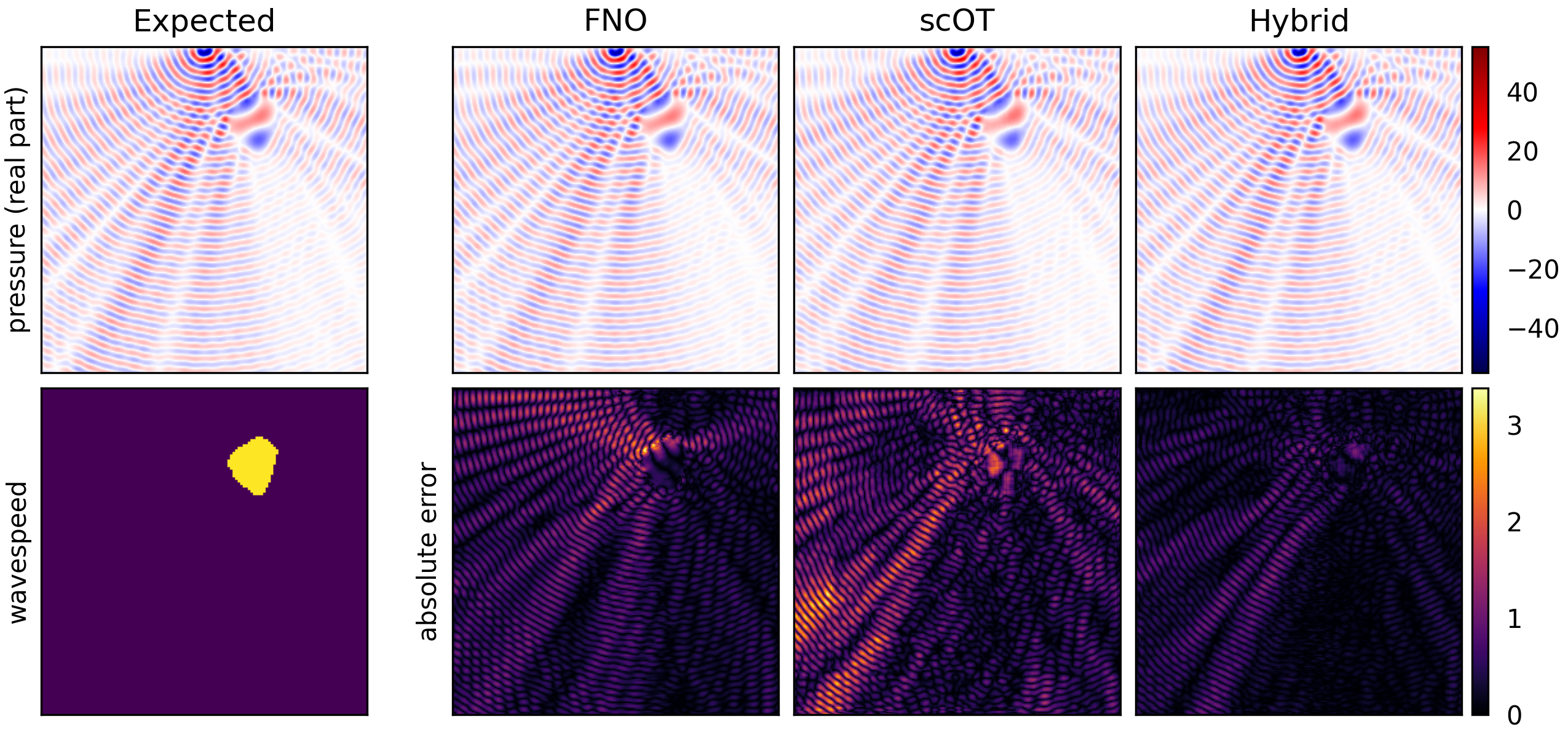}
\end{figure}

\begin{figure}[h!]
    \centering
    \includegraphics[width=0.8\linewidth]{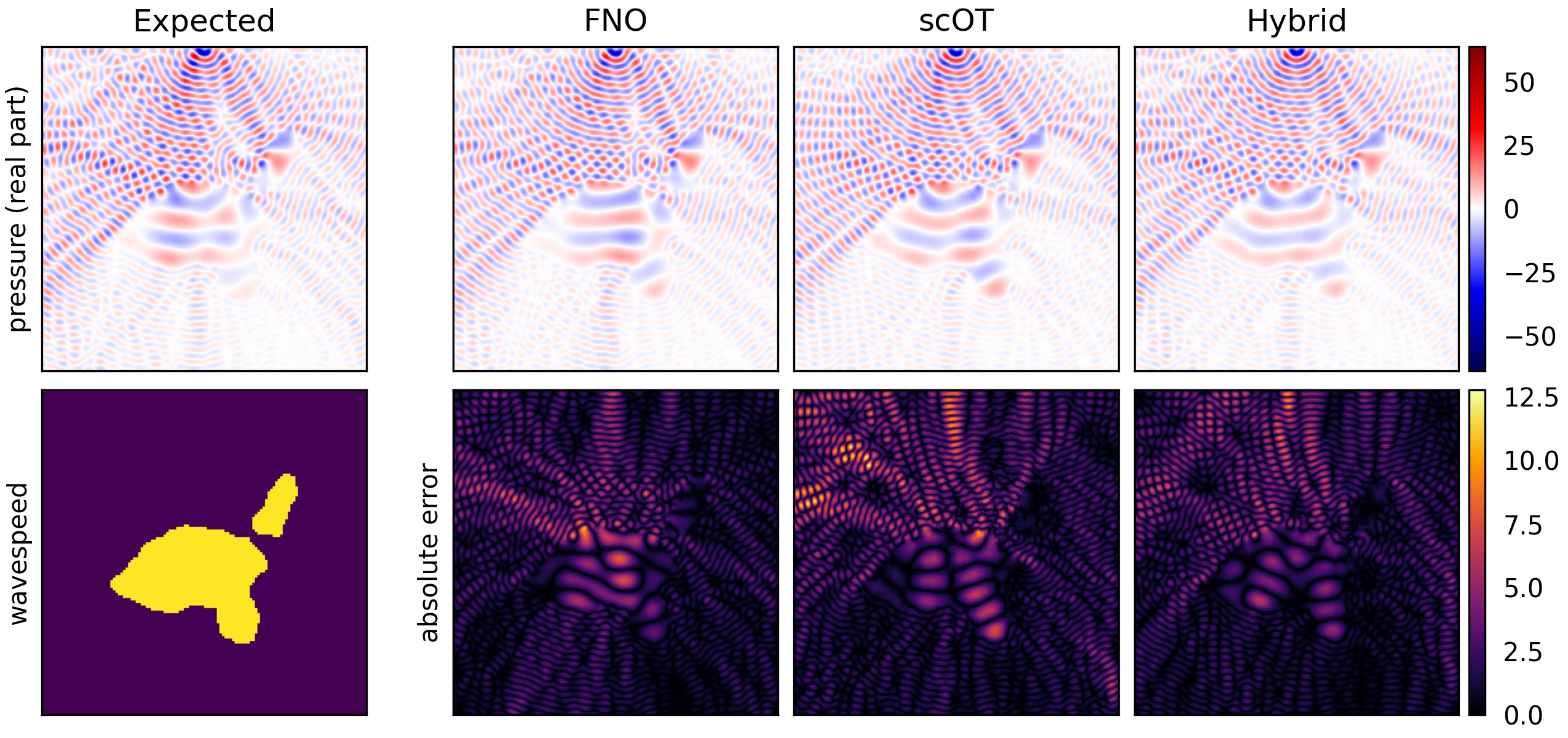}
\end{figure}

\clearpage

%% file: refs.bib
@misc{loshchilov2019decoupledweightdecayregularization,
      title={Decoupled Weight Decay Regularization}, 
      author={Ilya Loshchilov and Frank Hutter},
      year={2019},
      eprint={1711.05101},
      archivePrefix={arXiv},
      primaryClass={cs.LG},
      url={https://arxiv.org/abs/1711.05101}, 
}

@misc{loshchilov2017sgdrstochasticgradientdescent,
      title={SGDR: Stochastic Gradient Descent with Warm Restarts}, 
      author={Ilya Loshchilov and Frank Hutter},
      year={2017},
      eprint={1608.03983},
      archivePrefix={arXiv},
      primaryClass={cs.LG},
      url={https://arxiv.org/abs/1608.03983}, 
}

@article{Faucher2021, doi = {10.21105/joss.02699}, url = {https://doi.org/10.21105/joss.02699}, year = {2021}, publisher = {The Open Journal}, volume = {6}, number = {57}, pages = {2699}, author = {Faucher, Florian}, title = {`hawen`: time-harmonic wave modeling and inversion using hybridizable discontinuous Galerkin discretization}, journal = {Journal of Open Source Software} }

@article{kossaifi2025librarylearningneuraloperators,
   author    = {Jean Kossaifi and
                  Nikola Kovachki and
                  Zongyi Li and
                  David Pitt and
                  Miguel Liu-Schiaffini and
                  Valentin Duruisseaux and
                  Robert Joseph George and
                  Boris Bonev and
                  Kamyar Azizzadenesheli and
                  Julius Berner and
                  Anima Anandkumar},
   title     = {A Library for Learning Neural Operators},
   journal   = {arXiv preprint arXiv:2412.10354},
   year      = {2025},
}

@article{tarantola1984,
  author  = {Tarantola, Albert},
  title   = {Inversion of seismic reflection data in the acoustic approximation},
  journal = {Geophysics},
  year    = {1984},
  volume  = {49},
  number  = {8},
  pages   = {1259--1266},
  doi     = {10.1190/1.1441754}
}

@article{pratt1999,
  author  = {Pratt, R. Gerhard},
  title   = {Seismic waveform inversion in the frequency domain, Part 1: Theory and verification in a physical scale model},
  journal = {Geophysics},
  year    = {1999},
  volume  = {64},
  number  = {3},
  pages   = {888--901},
  doi     = {10.1190/1.1444597}
}

@article{li2020fourier,
  title   = {Fourier Neural Operator for Parametric Partial Differential Equations},
  author  = {Li, Zongyi and Kovachki, Nikola B. and Azizzadenesheli, Kamyar and Liu, Burigede and Bhattacharya, Kaushik and Stuart, Andrew M. and Anandkumar, Anima},
  journal = {arXiv preprint arXiv:2010.08895},
  year    = {2020},
  url     = {https://arxiv.org/abs/2010.08895}
}

@article{lumley2001,
  author  = {Lumley, David},
  title   = {Time-lapse seismic reservoir monitoring},
  journal = {Geophysics},
  year    = {2001},
  volume  = {66},
  number  = {1},
  pages   = {50--53},
  doi     = {10.1190/1.1444921}
}

@incollection{arts2004sleipner,
  author    = {Arts, Rob and Eiken, Ola and Chadwick, Andy and Zweigel, Peter and van der Meer, Leon and Zinszner, Bernard},
  title     = {Seismic monitoring at the Sleipner underground CO$_2$ storage site (North Sea)},
  booktitle = {Geological Storage of Carbon Dioxide},
  series    = {Geological Society, London, Special Publications},
  volume    = {233},
  year      = {2004},
  pages     = {181--191},
  doi       = {10.1144/GSL.SP.2004.233.01.12},
  publisher = {Geological Society of London}
}

@article{virieux2009overview,
  author  = {Virieux, Jean and Operto, St{\'e}phane},
  title   = {An overview of full-waveform inversion in exploration geophysics},
  journal = {Geophysics},
  year    = {2009},
  volume  = {74},
  number  = {6},
  pages   = {WCC1--WCC26},
  doi     = {10.1190/1.3238367}
}

@misc{deHoop2024universal-in-context,
      title={Transformers are Universal In-context Learners}, 
      author={Takashi Furuya and Maarten V. de Hoop and Gabriel Peyré},
      year={2024},
      eprint={2408.01367},
      archivePrefix={arXiv},
      primaryClass={cs.CL},
      url={https://arxiv.org/abs/2408.01367}, 
}

@article{lara_benitez2024ood,
  title   = {Out-of-distributional risk bounds for neural operators with applications to the Helmholtz equation},
  author  = {Lara Benitez, Jos{\'e} Antonio and Furuya, Takashi and Faucher, Florian and Kratsios, Anastasis and Tricoche, Xavier and de Hoop, Maarten V.},
  journal = {Journal of Computational Physics},
  year    = {2024},
  volume  = {513},
  pages   = {113168},
  doi     = {10.1016/j.jcp.2024.113168}
}

@article{kratsios2024mixture,
  title={Mixture of experts soften the curse of dimensionality in operator learning},
  author={Kratsios, Anastasis and Furuya, Takashi and Benitez, Jose Antonio Lara and Lassas, Matti and de Hoop, Maarten},
  journal={arXiv preprint arXiv:2404.09101},
  year={2024}
}

@article{liu2024wavebench,
  title={Wavebench: Benchmarking data-driven solvers for linear wave propagation pdes},
  author={Liu, Tianlin and Benitez, Jose Antonio Lara and Faucher, Florian and Khorashadizadeh, Amirehsan and de Hoop, Maarten V and Dokmani{\'c}, Ivan},
  journal={Transactions on Machine Learning Research Journal},
  year={2024}
}

@book{rasmussen2006gp,
  title={Gaussian Processes for Machine Learning},
  author={Rasmussen, Carl Edward and Williams, Christopher K. I.},
  year={2006},
  publisher={MIT Press},
  isbn={9780262182539}
}

@INPROCEEDINGS{Swin-v2,
  author={Liu, Ze and Hu, Han and Lin, Yutong and Yao, Zhuliang and Xie, Zhenda and Wei, Yixuan and Ning, Jia and Cao, Yue and Zhang, Zheng and Dong, Li and Wei, Furu and Guo, Baining},
  booktitle={2022 IEEE/CVF Conference on Computer Vision and Pattern Recognition (CVPR)}, 
  title={Swin Transformer V2: Scaling Up Capacity and Resolution}, 
  year={2022},
  volume={},
  number={},
  pages={11999-12009},
  doi={10.1109/CVPR52688.2022.01170}}

@inproceedings{Hao2024DPOT,
  author    = {Hao, Zhiyan and Su, Chang and Liu, Shuxiao and Berner, Jonas and Ying, Chengrun and Su, Hang and Anandkumar, Anima and Song, Jiaming and Zhu, Jun},
  title     = {DPOT: Auto-regressive denoising operator transformer for large-scale {PDE} pre-training},
  booktitle = {Proceedings of the 41st International Conference on Machine Learning},
  year      = {2024},
  pages     = {17616--17635},
  publisher = {PMLR}
}

@inproceedings{Herde2024Poseidon,
  author    = {Herde, Marcus and Raoni{\'c}, Bojan and Rohner, Thomas and K{\"a}ppeli, Rolf and Molinaro, Roberto and de B{\'e}zenac, Emmanuel and Mishra, Siddhartha},
  title     = {Poseidon: Efficient foundation models for {PDEs}},
  booktitle = {Advances in Neural Information Processing Systems},
  year      = {2024}
}

@article{Li2022TransformerPDE,
  author    = {Li, Zijie and Meidani, Hadi and Farimani, Amir Barati},
  title     = {Transformer for partial differential equations' operator learning},
  journal   = {arXiv preprint arXiv:2205.13671},
  year      = {2022},
  eprint    = {2205.13671},
  archivePrefix = {arXiv}
}

@inproceedings{McCabe2024MultiplePhysics,
  author    = {McCabe, Matthew and R{\'e}galdo-Saint Blancard, Bastien and Parker, Laura H. and Ohana, Roy and Cranmer, Miles and Bietti, Alberto and Eickenberg, Michael and Golkar, Siavash and Krawezik, G{\'e}rard and Lanusse, Fran{\c{c}}ois and Pettee, Mark and Tesileanu, Tiberiu and Cho, Kyunghyun and Ho, Shirley},
  title     = {Multiple physics pretraining for spatiotemporal surrogate models},
  booktitle = {Advances in Neural Information Processing Systems},
  year      = {2024}
}

@techreport{Chadwick2009BestPractice,
  author      = {Chadwick, Andrew and Arts, Rob and Bernstone, Christian and May, Felix and Thibeau, Sylvain and Zweigel, Peter},
  title       = {Best practice for the storage of {CO$_2$} in saline aquifers: observations and guidelines from the {SACS} and {CO$_2$STORE} projects},
  institution = {British Geological Survey},
  year        = {2009}
}

@article{Chen2024NSNO,
  author  = {Chen, Fang and Liu, Zhi and Lin, Guang and Chen, Jinjun and Shi, Zhong},
  title   = {NSNO: Neumann Series Neural Operator for Solving Helmholtz Equations in Inhomogeneous Medium},
  journal = {Journal of Systems Science and Complexity},
  volume  = {37},
  number  = {2},
  pages   = {413--440},
  year    = {2024}
}

@article{Cheng2021qPPropagators,
  author  = {Cheng, Jing and Alkhalifah, Tariq and Wu, Zhen and Zou, Puning and Wang, Chunlin},
  title   = {Simulating propagation of separated wave modes in general anisotropic media, Part {I}: qP-wave propagators},
  journal = {Geophysics},
  volume  = {86},
  number  = {1},
  pages   = {C1--C17},
  year    = {2021}
}

@article{Clayton1981BornWKBJ,
  author  = {Clayton, Robert W. and Stolt, Robert H.},
  title   = {A Born-{WKBJ} inversion method for acoustic reflection data},
  journal = {Geophysics},
  volume  = {46},
  number  = {11},
  pages   = {1559--1567},
  year    = {1981}
}

@article{deHoop2000WavefieldReciprocity,
  author  = {de Hoop, Maarten V. and de Hoop, Adriaan T.},
  title   = {Wavefield reciprocity and optimization in remote sensing},
  journal = {Proceedings of the Royal Society of London. Series A: Mathematical, Physical and Engineering Sciences},
  volume  = {456},
  number  = {1996},
  pages   = {641--682},
  year    = {2000}
}

@book{Fichtner2011Book,
  author    = {Fichtner, Andreas},
  title     = {Full seismic waveform modelling and inversion},
  publisher = {Springer},
  year      = {2011}
}

@article{Fichtner2006AdjointTheory,
  author  = {Fichtner, Andreas and Bunge, Hans-Peter and Igel, Heiner},
  title   = {The adjoint method in seismology: {I}. Theory},
  journal = {Physics of the Earth and Planetary Interiors},
  volume  = {157},
  number  = {1--2},
  pages   = {86--104},
  year    = {2006}
}

@article{Huang2024PINNsFormer,
  author  = {Huang, Xinyu and Alkhalifah, Tariq},
  title   = {PINNsFormer: A transformer-based framework for physics-informed neural networks},
  journal = {arXiv preprint arXiv:2307.11833},
  year    = {2024},
  eprint  = {2307.11833},
  archivePrefix = {arXiv}
}

@article{Koketsu2004VoxelFEM,
  author  = {Koketsu, Kazuo and Fujiwara, Hiroyuki and Ikegami, Yoshio},
  title   = {Finite-element simulation of seismic ground motion with a voxel mesh},
  journal = {Pure and Applied Geophysics},
  volume  = {161},
  number  = {11},
  pages   = {2183--2198},
  year    = {2004}
}

@article{Kovachki2021FNOTheory,
  author  = {Kovachki, Nikola B. and Lanthaler, Samuel and Mishra, Siddhartha},
  title   = {On universal approximation and error bounds for Fourier neural operators},
  journal = {Journal of Machine Learning Research},
  volume  = {22},
  number  = {290},
  pages   = {1--76},
  year    = {2021}
}

@inproceedings{Krishnapriyan2021PINNFailureModes,
  author    = {Krishnapriyan, Aditi and Gholami, Amir and Zhe, Shi and Kirby, Robert and Mahoney, Michael W.},
  title     = {Characterizing possible failure modes in physics-informed neural networks},
  booktitle = {Advances in Neural Information Processing Systems},
  volume    = {34},
  pages     = {26548--26560},
  year      = {2021}
}

@article{Marfurt1984FDvsFEMAccuracy,
  author  = {Marfurt, Kurt J.},
  title   = {Accuracy of finite-difference and finite-element modeling of the scalar and elastic wave equations},
  journal = {Geophysics},
  volume  = {49},
  number  = {5},
  pages   = {533--549},
  year    = {1984}
}

@article{Moczo2002HeterogeneousFD,
  author  = {Moczo, Peter and Kristek, Jaroslav and Vavry{\v{c}}uk, V{\'a}clav and Archuleta, Ralph J. and Halada, Lubom{\'\i}r},
  title   = {3D heterogeneous staggered-grid finite-difference modeling of seismic motion with volume harmonic and arithmetic averaging of elastic moduli and densities},
  journal = {Bulletin of the Seismological Society of America},
  volume  = {92},
  number  = {8},
  pages   = {3042--3066},
  year    = {2002}
}

@article{Padovani1994FEMSeismic,
  author  = {Padovani, Enrico and Priolo, Enrico and Seriani, Giovanni},
  title   = {Low- and high-order finite element method: Experience in seismic modeling},
  journal = {Journal of Computational Acoustics},
  volume  = {2},
  number  = {4},
  pages   = {371--422},
  year    = {1994}
}

@article{Pratt1990CrossholeTomography,
  author  = {Pratt, R. G. and Worthington, M. H.},
  title   = {Inverse theory applied to multi-source cross-hole tomography. Part 1: Acoustic wave-equation method},
  journal = {Geophysical Prospecting},
  volume  = {38},
  number  = {3},
  pages   = {287--310},
  year    = {1990}
}

@article{Raissi2019PINNs,
  author  = {Raissi, Maziar and Perdikaris, Paris and Karniadakis, George Em},
  title   = {Physics-informed neural networks: A deep learning framework for solving forward and inverse problems involving nonlinear partial differential equations},
  journal = {Journal of Computational Physics},
  volume  = {378},
  pages   = {686--707},
  year    = {2019}
}

@article{Robertsson1994ViscoelasticFD,
  author  = {Robertsson, Johan O. and Blanch, Joachim O. and Symes, William W.},
  title   = {Viscoelastic finite-difference modeling},
  journal = {Geophysics},
  volume  = {59},
  number  = {9},
  pages   = {1444--1456},
  year    = {1994}
}

@article{Schmelzbach2016GeothermalImaging,
  author  = {Schmelzbach, C. and Greenhalgh, S. and Reiser, F. and Girard, J.-F. and Bretaudeau, F. and Capar, L. and Bitri, A.},
  title   = {Advanced seismic processing/imaging techniques and their potential for geothermal exploration},
  journal = {Interpretation},
  volume  = {4},
  number  = {4},
  pages   = {SR1--SR18},
  year    = {2016}
}

@book{Sheriff1995ExplorationSeismology,
  author    = {Sheriff, Robert E. and Geldart, Lloyd P.},
  title     = {Exploration Seismology},
  publisher = {Cambridge University Press},
  year      = {1995}
}

@article{Song2022PINNsVTI,
  author  = {Song, Chen and Alkhalifah, Tariq and Waheed, Umair Bin},
  title   = {Solving the frequency-domain acoustic {VTI} wave equation using physics-informed neural networks},
  journal = {Geophysical Journal International},
  volume  = {229},
  number  = {2},
  pages   = {846--859},
  year    = {2022}
}

@article{Tromp2005BananaDoughnut,
  author  = {Tromp, Jeroen and Tape, Carl and Liu, Qin},
  title   = {Seismic tomography, adjoint methods, time reversal and banana-doughnut kernels},
  journal = {Geophysical Journal International},
  volume  = {160},
  number  = {1},
  pages   = {195--216},
  year    = {2005}
}

@article{Virieux1984SH,
  author  = {Virieux, Jean},
  title   = {SH-wave propagation in heterogeneous media: Velocity-stress finite-difference method},
  journal = {Geophysics},
  volume  = {49},
  number  = {11},
  pages   = {1933--1942},
  year    = {1984}
}

@article{Wang2021WhenPINNsFail,
  author  = {Wang, Sifan and Yu, Xuhui and Perdikaris, Paris},
  title   = {When and why {PINNs} fail to train: A neural tangent kernel perspective},
  journal = {Journal of Computational Physics},
  volume  = {449},
  pages   = {110768},
  year    = {2021}
}

@book{Yilmaz2001SeismicDataAnalysis,
  author    = {Yilmaz, {\"O}z},
  title     = {Seismic Data Analysis: Processing, Inversion, and Interpretation of Seismic Data},
  publisher = {Society of Exploration Geophysicists},
  year      = {2001}
}

@inproceedings{Yin2023ImplicitForecasting,
  author    = {Yin, Yujia and Kirchmeyer, Matthieu and Franceschi, Jean-Yves and Rakotomamonjy, Alain and Gallinari, Patrick},
  title     = {Continuous {PDE} dynamics forecasting with implicit neural representations},
  booktitle = {International Conference on Learning Representations},
  year      = {2023}
}

@article{Zhu2023FourierDeepONet,
  author  = {Zhu, Mengling and Feng, Shi and Lin, Yubin and Lu, Lu},
  title   = {Fourier-{DeepONet}: Fourier-enhanced deep operator networks for full waveform inversion with improved accuracy, generalizability, and robustness},
  journal = {Computer Methods in Applied Mechanics and Engineering},
  volume  = {416},
  pages   = {116300},
  year    = {2023}
}

@article{cheng2025seismicgno,
  author  = {Cheng, Shijun and Taufik, Mohammad H. and Alkhalifah, Tariq},
  title   = {Seismic wavefield solutions via physics-guided generative neural operator},
  journal = {arXiv preprint arXiv:2503.06488},
  year    = {2025},
  doi     = {10.48550/arXiv.2503.06488},
  url     = {https://arxiv.org/abs/2503.06488}
}

@unpublished{florian2025,
  TITLE = {{Enriching continuous Lagrange finite element approximation spaces using neural networks}},
  AUTHOR = {Barucq, H{\'e}l{\`e}ne and Duprez, Michel and Faucher, Florian and Franck, Emmanuel and Lecourtier, Fr{\'e}d{\'e}rique and Lleras, Vanessa and Michel-Dansac, Victor and Victorion, Nicolas},
  URL = {https://hal.science/hal-04935072},
  NOTE = {working paper or preprint},
  YEAR = {2025},
  MONTH = Feb,
  HAL_ID = {hal-04935072},
  HAL_VERSION = {v3},
}

@misc{wen2026geometryawareoperatortransformer,
      title={Geometry Aware Operator Transformer as an Efficient and Accurate Neural Surrogate for PDEs on Arbitrary Domains}, 
      author={Shizheng Wen and Arsh Kumbhat and Levi Lingsch and Sepehr Mousavi and Yizhou Zhao and Praveen Chandrashekar and Siddhartha Mishra},
      year={2026},
      eprint={2505.18781},
      archivePrefix={arXiv},
      primaryClass={cs.LG},
      url={https://arxiv.org/abs/2505.18781}, 
}

@article{Guibas2021AdaptiveFN,
  title={Adaptive Fourier Neural Operators: Efficient Token Mixers for Transformers},
  author={John Guibas and Morteza Mardani and Zong-Yi Li and Andrew Tao and Anima Anandkumar and Bryan Catanzaro},
  journal={ArXiv},
  year={2021},
  volume={abs/2111.13587},
  url={https://api.semanticscholar.org/CorpusID:244709538}
}

@article{NO-Stuart-Kovachki-Anandkumar-et-al,
author = {Kovachki, Nikola and Li, Zongyi and Liu, Burigede and Azizzadenesheli, Kamyar and Bhattacharya, Kaushik and Stuart, Andrew and Anandkumar, Anima},
title = {Neural operator: learning maps between function spaces with applications to PDEs},
year = {2023},
issue_date = {January 2023},
publisher = {JMLR.org},
volume = {24},
number = {1},
issn = {1532-4435},
journal = {J. Mach. Learn. Res.},
month = jan,
articleno = {89},
numpages = {97}
}

@article{zou2024deep,
  title={U-NO: U-shaped neural operators for solving the Helmholtz equation in the frequency domain},
  author={Zou, Zongyi and Rahman, M. A. and Ross, Zachary E and Azizzadenesheli, Kamyar},
  journal={Transactions on Machine Learning Research},
  year={2024}
}

@article{wang2024transfer,
  title={Transfer learning of neural operators for seismic wavefield prediction across different source locations and frequencies},
  author={Wang, H and Alkhalifah, Tariq and Huang, Xuqing},
  journal={Geophysical Journal International},
  volume={236},
  number={3},
  pages={1567--1580},
  year={2024},
  publisher={Oxford University Press}
}

@article{ma2025picno,
  title={PICNO: Physics-informed convolutional neural operators for seismic wavefield modeling},
  author={Ma, Chao and others},
  journal={Geophysics},
  note={In Press},
  year={2025},
  publisher={Society of Exploration Geophysicists}
}

@article{song2024traveltime,
  title={Solving the eikonal equation for traveltime computation using physics-informed neural operators},
  author={Song, Chao and Alkhalifah, Tariq and Waheed, Umair bin},
  journal={IEEE Transactions on Geoscience and Remote Sensing},
  volume={62},
  year={2024},
  publisher={IEEE}
}

@article{cheng2025gno,
  title={Seismic wavefield solutions via physics-guided generative neural operator},
  author={Cheng, S and Taufik, M. H. and Alkhalifah, Tariq},
  journal={arXiv preprint arXiv:2503.06488},
  year={2025}
}

@article{lehmann2024ffno,
  title={3D elastic wave propagation with a factorized Fourier neural operator (F-FNO)},
  author={Lehmann, F and Gatti, F and Bertin, M and Clouteau, D},
  journal={Computer Methods in Applied Mechanics and Engineering},
  volume={420},
  pages={116718},
  year={2024},
  publisher={Elsevier}
}

@inproceedings{qin2024spectral,
  title={Spectral bias in neural operators: Analysis and mitigation strategies},
  author={Qin, Tong and others},
  booktitle={Proceedings of the 38th Conference on Neural Information Processing Systems (NeurIPS)},
  year={2024}
}

@article{huang2025physics,
  title={Physics-informed neural operator for seismic inversion: Addressing the generalization gap in smooth perturbed models},
  author={Huang, Xuqing and Alkhalifah, Tariq},
  journal={Geophysics},
  volume={90},
  number={1},
  year={2025},
  publisher={Society of Exploration Geophysicists}
}

@inproceedings{Dong2023GNOT,
author = {Hao, Zhongkai and Wang, Zhengyi and Su, Hang and Ying, Chengyang and Dong, Yinpeng and Liu, Songming and Cheng, Ze and Song, Jian and Zhu, Jun},
title = {GNOT: a general neural operator transformer for operator learning},
year = {2023},
publisher = {JMLR.org},
booktitle = {Proceedings of the 40th International Conference on Machine Learning},
articleno = {509},
numpages = {14},
location = {Honolulu, Hawaii, USA},
series = {ICML'23}
}

@article{Alkhalifah2024-scattered-residual,
    author = {Huang, Xinquan and Alkhalifah, Tariq},
    title = {Learned frequency-domain scattered wavefield solutions using neural operators},
    journal = {Geophysical Journal International},
    volume = {241},
    number = {3},
    pages = {1467-1478},
    year = {2025},
    month = {03},
    issn = {1365-246X},
    doi = {10.1093/gji/ggaf113},
    url = {https://doi.org/10.1093/gji/ggaf113},
    eprint = {https://academic.oup.com/gji/article-pdf/241/3/1467/62750062/ggaf113.pdf},
}

@article{HUANG2024-LordNet,
title = {LordNet: An efficient neural network for learning to solve parametric partial differential equations without simulated data},
journal = {Neural Networks},
volume = {176},
pages = {106354},
year = {2024},
issn = {0893-6080},
doi = {https://doi.org/10.1016/j.neunet.2024.106354},
url = {https://www.sciencedirect.com/science/article/pii/S0893608024002788},
author = {Xinquan Huang and Wenlei Shi and Xiaotian Gao and Xinran Wei and Jia Zhang and Jiang Bian and Mao Yang and Tie-Yan Liu}
}

@inproceedings{raonic2023convolutional,
  title={Convolutional Neural Operators for robust and accurate learning of PDEs},
  author={Raoni{\'c}, Bogdan and Molinaro, Roberto and De Ryck, Tim and Rohner, Tobias and Bartolucci, Francesca and Alaifari, Rima and Mishra, Siddhartha and de B{\'e}zenac, Emmanuel},
  booktitle={Advances in Neural Information Processing Systems},
  volume={36},
  pages={52658--52695},
  year={2023}
}
